\documentclass[preprint,showpacs,amsmath,amssymb,pre,superscriptaddress]{revtex4}

\usepackage{graphicx}
\usepackage{dcolumn}
\usepackage{bm}

\begin{document}

\title{Classical Dynamics of the Time-Dependent Elliptical Billiard}

\date{\today}

\pacs{05.45.-a,05.45.Ac,05.45.Pq}

\author{Florian Lenz}

\email[]{lenz@physi.uni-heidelberg.de}

\affiliation{Physikalisches Institut, Universit\"at Heidelberg, Philosophenweg 12, 69120 Heidelberg, Germany}%
\author{Fotis K. Diakonos}
\affiliation{Department of Physics, University of Athens, GR-15771 Athens, Greece}
\author{Peter Schmelcher}


\affiliation{Physikalisches Institut, Universit\"at Heidelberg, Philosophenweg 12, 69120 Heidelberg, Germany}%

\affiliation{Theoretische Chemie, Institut f\"ur Physikalische Chemie, Universit\"at Heidelberg, INF 229, 69120 Heidelberg, Germany}%

\date{\today}


\begin{abstract}\label{ch:abstract}

In this work we study the nonlinear dynamics of the static
and the driven ellipse.  In the static case, we
find numerically an asymptotical algebraic decay  for the escape of an
ensemble of non-interacting particles
through a small hole due to the integrable structure of the phase space of the
system. Furthermore, for a certain hole position a saturation value in the decay
 that can
be tuned arbitrarily by varying the eccentricity of the ellipse is observed and
explained. When
applying harmonic boundary oscillations this saturation value
caused by librator type orbits is gradually destroyed via two fundamental
processes which are discussed in detail. As a
result, an amplitude dependent emission rate is obtained in the long time
behavior of the decay, suggesting that the driven elliptical billiard can be
used as a controllable source of particles.

\end{abstract}

\maketitle

\section{Introduction}\label{sec:introduction}

Billiards belong to the most widely studied Hamiltonian
systems. They possess many  classical and quantum  mechanical
properties of complex dynamical systems \cite{Ber81,Sto99,Gut90}.
Moreover, models of statistical mechanics can be reduced to billiards
\cite{Ego00}. For example, one of the simplest billiards,
particles inside a rectangular box, is an idealization of the
physical situation of nucleons confined inside a nucleus
\cite{Bau90}. Mathematically rigorous studies of billiards go back
to the early seventies, e.g. Bunimovich proved that  stadia are
ergodic \cite{Bun79}, using concepts developed by Sinai. In recent
years, a renewed interest in billiards has come up, due to the
possibilities of realizing them experimentally, for example by
using ultracold atoms confined in a laser potential \cite{Rai01},
microwave billiards \cite{Sto99, Gra92, Ste92}, or mesoscopic
quantum dots \cite{Mar92}.  Even for the design of directional
micro-lasers, billiards are relevant \cite{Noc97}. Besides this,
interesting theoretical results were obtained, including a
justification of a probabilistic approach to statistical mechanics
\cite{Zas99, Ego00}. Very recently, it has been shown
\cite{Bun05}, that a connection exists between billiards and one
of the major unsolved problems in mathematics, the Riemann
hypothesis: the authors found an analytic expression for the
escape rate of a circular billiard with two holes, involving a sum
over the zeros of the Riemann zeta function.

A natural generalization of billiards with a static boundary is to
apply a driving law to the billiard wall.  For instance, `Bohr's
liquid drop model' from nuclear physics can be regarded as a
time-dependent billiard \cite{Pai91}. For this simple looking
model still many  questions remain open \cite{Aba92}.  Another
example is in plasma physics, where time-dependent billiards
represent models for acceleration of particles in a magnetic
bottle, see \cite{Koi95} and ref. therein. In conclusion, there
are many branches of physics in which billiards, specifically
time-dependent billiards, serve as  models for more complex
systems, capturing the key features and behavior of the original
problem.

Ultracold atoms in a billiard formed by beams of light allow for
the possibility of generating arbitrary geometries and changing
them in time, as well as varying parameters such as beam width,
softness of the potential etc. in time. Of special interest is the
possibility to probe the dynamics by analyzing the escape rates
\cite{Dav00,Dav012,Dav011,Rai01,Dav02}, which has up to date only been performed
for static billiards. Introducing noise and
decoherence and studying the role of quantum and many-body effects
are further intriguing goals \cite{Rai01}.

Regarding time-dependent billiards, there exist several investigations in the literature
\cite{LL92, Los99, Los00, Los02,Koi95b, Koi95, Iti03,Sch06,Coh03}.
A crucial question for these systems is whether Fermi acceleration
occurs or not. This is examined in refs. \cite{LL92, Los99, Los00,
Los02} and very recently in ref. \cite{Sch06}. In \cite{LL92} it
is shown that when using smooth forcing functions, the existence
of invariant spanning KAM curves in phase space limits the energy
gain of the particles, whereas non-smooth forcing functions,
especially random oscillation lead to unbounded energy gain, see also
ref. \cite{Sch06} and references therein. In ref. \cite{Los99},
the authors conclude with the hypothesis:  \emph{``A random
element in a billiard with a fixed boundary is a sufficient
condition for the Fermi acceleration in the system when a boundary
perturbation is introduced.'' }.

Within the existing studies of classical time-dependent billiards
only little emphasis is put on systems with a finite horizon and
(to our knowledge) none to the corresponding escape rates. Very
few works deal with the time-dependent ellipse \cite{Koi95b,
Koi95, Iti03}. In ref. \cite{Koi95}, the average velocity as a
function of time and the Poincar\'e surface of section of the
dynamics of the ellipse for different driving laws are studied
numerically. Depending on the driving laws and the initial
velocity of an ensemble, the integrable structure of the phase
space is more or less destroyed compared to the static case. The
velocity of the particles stays bounded in all cases, i.e. no
Fermi acceleration occurs. The authors point out, that all
conditions are satisfied in order to apply Douady's theorem
\cite{Dou82} which predicts this boundedness of the velocity. A
mathematical study of periodically driven ellipses is given in
\cite{Koi95b}. The authors show that in principle it is possible
to destroy the diametral 2-periodic orbit via boundary
oscillations and give strong evidence that the opposite -
stabilizing an unstable periodic orbit with the use of driving -
is not possible.

The above discussion shows that essentially little is known about escape rates in
classical time-dependent billiards. Apart from being of
fundamental interest this type of driven dynamical systems is
nowadays well within the reach of experiments, as indicated above. Moreover our
investigation will demonstrate that
time-dependent billiards might provide us with a tunable source
of particles. As we shall see, the escape rate and the
velocity distribution of the escaping particles strongly depends
on the driving properties, such as amplitude and frequency of a
periodic driving.

In this work, which is an extension of our recently published letter \cite{Le07}, we focus on the driven elliptical billiard. Its static counterpart is integrable, due to the existence of a second constant of motion, the product of the angular momenta around the foci.  Thus the phase space of the ellipse  possesses  a more complex structure, consisting of librators and rotators, than the prototype integrable billiard, the circle. Naturally, it would be also interesting to study driven billiards whose static counterparts have mixed or chaotic dynamics. Yet the clear partition of the phase space into librators and rotators of the ellipse simplifies the analysis in the presence of the driving considerably. This allows us to study  e.g. transition between librator and rotator orbits and  to discuss associated physical phenomena in an intuitive way.

This article is structured as follows: In  section
\ref{ch:StaticEllipse} we discuss fundamental properties and
escape rates in the static ellipse. The generalization to
time-dependent ellipses is treated in section
\ref{ch:TimeDepEllipse}. The fundamental processes leading to the
destruction of the librator orbits are displayed in section
\ref{ch:DesTypII} followed by  an analysis of the angular
momentum, section \ref{ch:AngMom}, and the velocity, section
\ref{ch:Velocity}. Finally, a summary is given in the last
section.

\section{Static Ellipse}\label{ch:StaticEllipse}
\subsection{Fundamental Properties of the Dynamics in the
Ellipse}\label{ch:FundamentalProp}
In a two-dimensional static billiard, the  orbit of a particle can
be completely specified by providing the sequence of its positions
$s_i$ (measured by the arclength) or $\varphi_i$ (see eq.
\ref{eq:ElImplicit}) on the boundary $\mathcal{B}$  and the
directions $p_i = \cos \alpha_i$ immediately after each
collision, since the particles travel ballistically in between
collisions, where $\alpha_i$ is the angle between the forward
pointing tangent and the velocity of the particle at the $i$-th
collision point. The corresponding discrete mapping  $\mathcal{M}$
is area preserving in the phase space variables $s$ and $p$
\cite{Ber81}. The boundary $\mathcal{B}$ of an ellipse is given by
\begin{equation}\label{eq:ElImplicit}
   \mathcal{B} = \left\{ x(\varphi) = A  \cos \varphi,y(\varphi) = B  \sin
\varphi)| 0 \le \varphi <
    2\pi  \right\}
\end{equation}
with $A>B>0$, thus $A$ and $B$ being the long and  the
short half-diameter respectively. The dimensionless numerical
eccentricity can be written as $\varepsilon =
\sqrt{1-{B^2}/{A^2}}$.

In anticipation of the time-dependent problem, we describe the
direction of a particle by its velocity $\bm{v}=(v_x,v_y)$.  If we
demand without loss of generality $|\bm{v}|=1$,  there is a one to one correspondence between
the velocity $\bm{v}$ and $p$ at the collision points. At a certain time $t$, the position
of the particle starting at $t=0$ at $\bm{x}_0= (x_0,y_0) \in \mathcal{B}$ with
the velocity $(v_{x,0},v_{y,0})$
is given by
\begin{subequations}\label{eq:ParticleProp}
\begin{eqnarray}
  x(t) &=& x_0 + v_{x,0} t \\
  y(t) &=& y_0 + v_{y,0} t.
\end{eqnarray}
\end{subequations}
The particle will hit the boundary at $\bm{x}_1$ at the time
$t_1$.
\begin{equation}\label{eq:Time1}
    t_1 = -\frac{2B^2 x_0 v_{x,0} + 2A^2 y_0 v_{y,0}}{(A v_{y,0})^2 + (B v_{x,0})^2}
\end{equation}
To get the new velocity
$\bm{v}_1$, we parameterize  $\mathbf{x}_1 \in \mathcal{B}$ by
$\varphi_1$ and calculate the inward pointing normal vector
$\hat{\mathbf{n}}_1, \, |\hat{\mathbf{n}}_1|=1$ at $\varphi_1$. This results in
\begin{equation}\label{eq:NextVelocity}
    \mathbf{v}_1 = \mathbf{v}_0 -2 \left ( \hat{ \mathbf{n}}_1 \cdot \mathbf{v}_0 \right ) \cdot \hat{
    \mathbf{n}}_1.
\end{equation}
Equation \eqref{eq:NextVelocity} can be easily extended to
time-dependent boundaries, see section \ref{ch:TimeDepEllipse}, where momentum
transfer from the moving wall to the particle takes place.

\begin{figure}[ht]
\includegraphics[width=0.8\columnwidth]{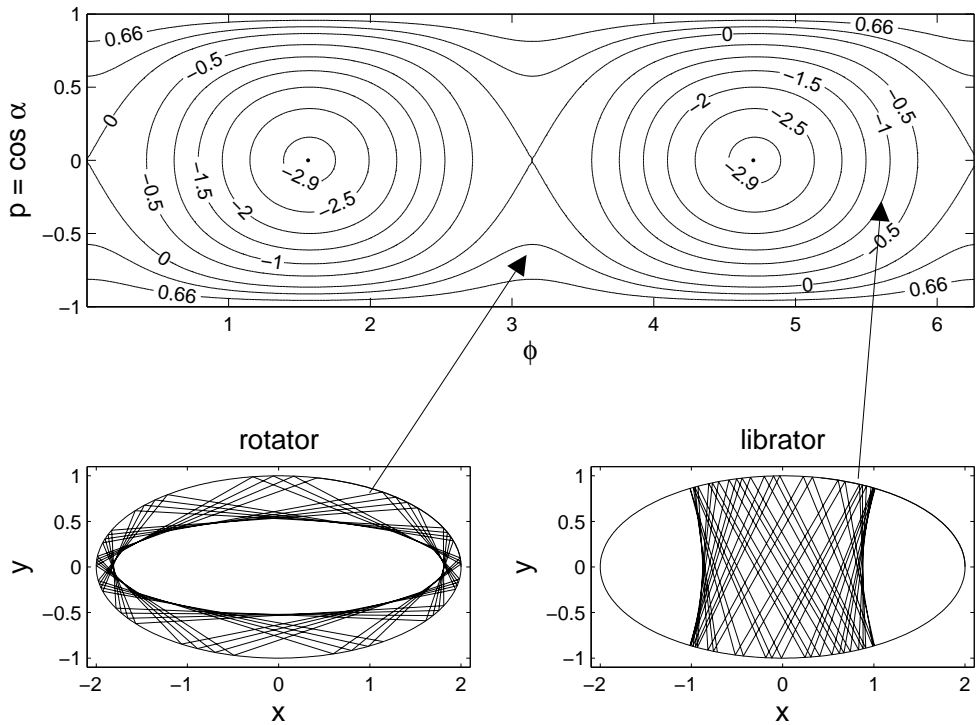}
\caption{PSS of the ellipse,
 (upper part) and typical trajectories (lower part), $A=2,\,B=1$. The rotator
orbit repeatedly touches a confocal ellipse, the librator orbit a confocal
hyperbola.}\label{fig:ElPSS}
\end{figure}

The dynamics in the ellipse is completely integrable, see Fig. \ref{fig:ElPSS}.
In
addition to the energy, there is another constant of motion
$F(\varphi,p)$, restricting the orbits to invariant curves in
phase space:

\begin{equation}\label{eq:El2ndIntPhi}
    F(\varphi,p)= \frac{p^2 (1+ (1-\varepsilon^2) \cot^2
\varphi)-\varepsilon^2}{1+ (1-\varepsilon^2) \cot^2 \varphi -
    \varepsilon^2}.
\end{equation}

$F(\varphi,p)$ can be interpreted as the product of the angular
momenta (PAM) about the two focus points \cite{Ber81}.
There are two different types of orbits, rotators and librators, in the
ellipse separated by the separatrix, see Fig. \ref{fig:ElPSS}.
Librators cross the x-axis between the two focus
points and touch repeatedly a confocal hyperbola. In the PSS, they
appear as deformed circles around elliptic fixed points, exploring
a limited range in $p$ as well as in $\varphi$. Rotator orbits
 travel around the ellipse, exploring every value of
$\varphi$, but  only a small range in $p$ (except if they are
very close to the separatrix), repeatedly touching a confocal
ellipse.

In terms of $F(p,\varphi)$, we can distinguish tree different
cases:
\begin{enumerate}
    \item $F(\varphi,p) > 0 $ corresponds to the rotator orbits with an
elliptical caustic.
    \item $F(\varphi,p) < 0 $ are the librators  with hyperbolic
caustic. This includes the two elliptic fixed points at $(\varphi=\pi/2,\,p=0)$
and $(\varphi=3\pi/2,\,p=0)$ corresponding to a period two orbit along the minor
axis with $F(\pi/2,0) = -\varepsilon^2/(1-\varepsilon^2)$.
    \item $F(\varphi,p) = 0 $ corresponds to the period two orbit along
the major axis with $(\varphi=0, \,p =
0)$ and $(\varphi=\pi, \,p = 0)$, seen in the PSS as two
hyperbolic fixed points.
\end{enumerate}

The topology of the PSS, Fig. \ref{fig:ElPSS}, is dominated by
two isolated periodic orbits. The condition for
stability is according to ref. \cite{Ber81}
\begin{equation}\label{eq:2BounceStab}
    \frac{\rho}{2R(\varphi)}-1 \begin{cases}
    >0   &  \text{unstable} \\
         <0 & \text{stable},
    \end{cases}
\end{equation}
where $R(\varphi)$ is the radius of curvature and $\rho$ is the distance in
coordinate space between two successive collisions. Using this
stability criterion, we get for the
periodic orbit along the long diameter $\rho/2R = 1/(1-\varepsilon^2)>1$
and therefore it is unstable. In contrast, the orbit along the short
diameter obeys $\rho/2R =  1-\varepsilon^2<1$ and is stable.

According to ref. \cite{Koi95}, Poncelet's theorem on projective
geometry can be applied to elliptical billiards \cite{Cha88}. It
states, that all trajectories possessing the same value of
$F(\varphi,p)$, share the same caustic and  the same type of dynamics.
In the case of periodic orbits this means, that given one periodic
orbit with a certain value of $F(\varphi,p)$, every trajectory
with the same value of $F(\varphi,p)$ is also periodic and has actually
the same period. Consequently, the only isolated periodic orbits
are the two discussed two-periodic orbits, all the other periodic
orbits are non-isolated and form families.

\subsection{Escape rates}\label{ch:StEscapeRate}
Let us focus now on the escape rates of a static elliptical
billiard with a hole placed on its boundary. In this subsection we
use for all simulations $A=2,\,B=1$, i.e. the numerical
eccentricity $\varepsilon = \sqrt3/2 \approx 0.87$. The number
$N_0$ of particles in the initial ensemble is $N_0= 10^7$. Each
particle is propagated  at most $10^6$ boundary collisions unless
it does not escape earlier. The initial conditions
$(\varphi_0,\alpha_0)_i,\,i=1,2,\dots 10^7$ (the index $i$ stands
for the $i$th particle) are chosen randomly. Note that the angle $\alpha_0$
is distributed uniformly in $[0,\,\pi]$, not $p_0=\cos \alpha_0$. We choose two
different hole positions $\varphi_\triangle=0$ and $\varphi_\triangle=\pi/2$. The hole size
$\triangle$ is set to $\triangle =0.03$ (measured in $\varphi$).
$\varphi_\triangle = 0$
corresponds to a hole lying in the very right of the ellipse of
Fig. \ref{fig:ElPSS}, and $\varphi_\triangle = \pi/2$ corresponds to the
location at
the very top of the ellipse. The
reason for this choice  is the following: If the hole lies at
$\varphi_\triangle= 0$, none of the librator orbits can escape, since
their invariant curves are not connected with the hole, whereas if
$\varphi_\triangle = \pi/2$, all orbits can participate in the decay.
In both cases, all rotator orbits can escape (as long as
they are not periodic), since they are ergodic with respect to the
phase space variable $\varphi$. The main data of these simulations
is the number of remaining particles in the billiard as a function
of the number of collisions  $N(n)$ or the elapsed time $N(t)$.
Note that we refer to $N(t)$ as the \textit{escape rate}, as done
in the literature, whereas  we will call $\dot{N}(t)$ the
\textit{emission rate}.

\begin{figure}[ht]
        \includegraphics[width=0.9\columnwidth]{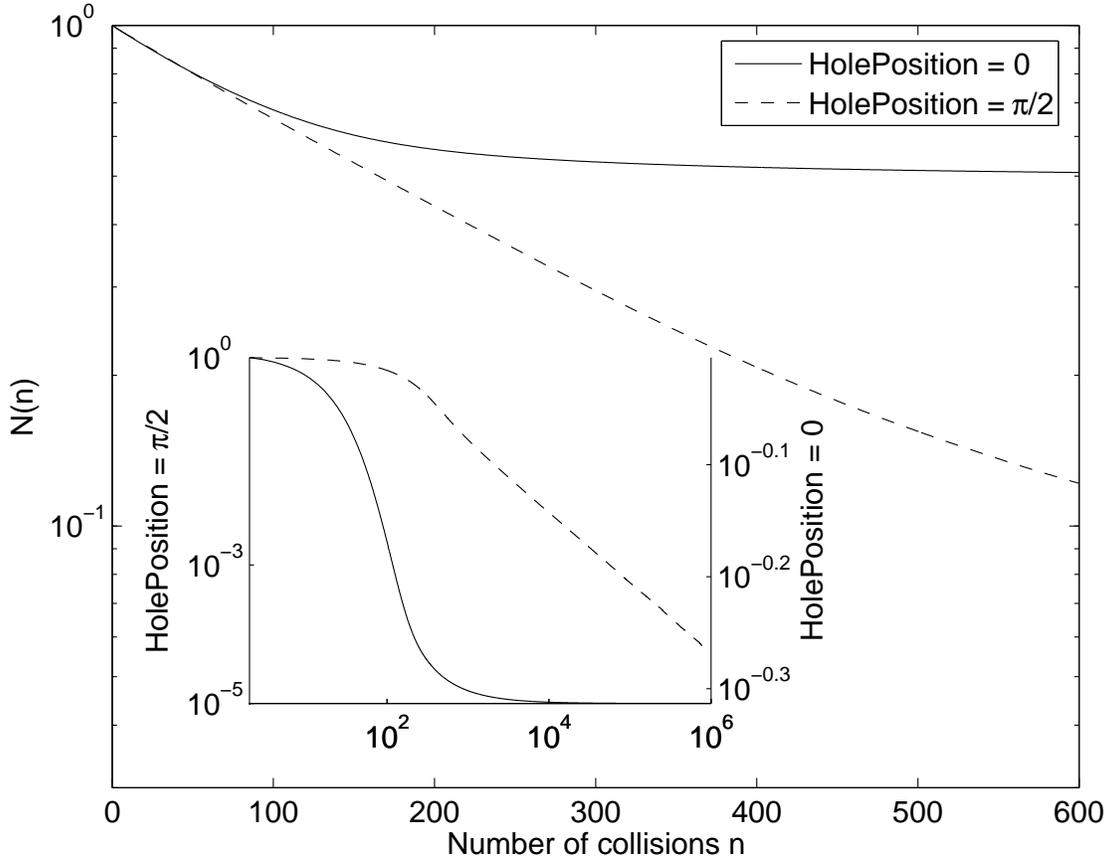}
        \caption{Semi-logarithmic plot of the escape rate, two different hole
positions are shown, double-logarithmic scale in the inset.}\label{fig:StEscSL}
\end{figure}

The results of the simulations are shown in Fig.
\ref{fig:StEscSL}. Two qualitatively
different behaviors of the decay are observed for the two
different hole positions:
\begin{enumerate}
    \item If $s_\triangle =\pi/2$, the fraction of remaining
    particles $N(n)$ as a function of the number of collisions
    approaches zero for $n\rightarrow \infty$.
    \item If $s_\triangle =0$, $N(n)$ approaches a saturation
    value $N_s(\varepsilon)>0$ after roughly $n=2\cdot 10^3$
    collisions.
\end{enumerate}
The saturation value $N_s(\varepsilon)$ in the case
$s_\triangle=0$ is of course caused by particles traveling on
librators. Since the librator orbits are not connected with
the hole, these particles will stay forever in the billiard. We
will derive an exact expression for $N_s(\varepsilon)$ in the next
section. In both cases of the hole position, the short time
behavior, Fig. \ref{fig:StEscSL}, of the decay is exponential
$N(n)\sim \exp (-\tau n)$ (roughly for the first 50 collisions in
the case $s_\triangle= 0$ and 300 collisions in the case
$s_\triangle= \pi/2$). The decay constant $\tau$ is approximately
given by $\tau \approx \triangle/2\pi$ \cite{Alt96}. The long time
behavior ($n>2\cdot 10^3$) of $N(n)$ in the case $s_\triangle=
\pi/2$ corresponds to an algebraic decay $N(n)\sim n^{-c}$, seen
as a straight line in the inset of Fig. \ref{fig:StEscSL}. This
power law decay is typical for integrable systems and known in the
literature, see e.g. ref. \cite{Bau90} or \cite{Mei86}, but there
is no work discussing the case of the ellipse, except for ref.
\cite{Dav011}, where the algebraic decay, even though not in such
detail, is observed experimentally.

A heuristic model explaining this algebraic behavior is provided in
\cite{Bau90}. The discussion given there holds for a
rectangular box, where  $|\bf{p\cdot e_n}|$ ($\bf{e_n}$ is the
unit vector normal to the opening) is a constant of motion.
Nevertheless, the results obtained there can be easily transferred
to the case of the ellipse by replacing $|\bf{p\cdot e_n}|$ by $F(\varphi,
p)$. According to \cite{Bau90}, the fraction of remaining
particles should decay for large $n$ like $ N(n)\sim n^{-1}$.
The extracted value from our data is $N(n)\sim
n^{-1.02}$ for $n>3\cdot 10^3$, i.e. in very good agreement with the analytical prediction.

\subsection{Saturation value $N_s(\varepsilon)$}\label{ch:SatValue}
Let us now study  whether the escape rates depend on the numerical
eccentricity $\varepsilon$. Indeed, the qualitative behavior of the decay
remains unchanged, only the saturation value $N_s(\varepsilon)$ is different
for different values of $\varepsilon$ for the hole at the short
side of the ellipse, $s_\triangle =0$. This becomes immediately
clear if one considers that $\varepsilon$ determines the
degree of deformation compared to the circle: Since in the circle
there are exclusively rotator orbits, $N_s(0)$ should be zero and
with increasing $\varepsilon$ the offset $N_s(\varepsilon)$ should
increase too.

\begin{figure}[ht]
        \includegraphics[width=0.9\columnwidth]{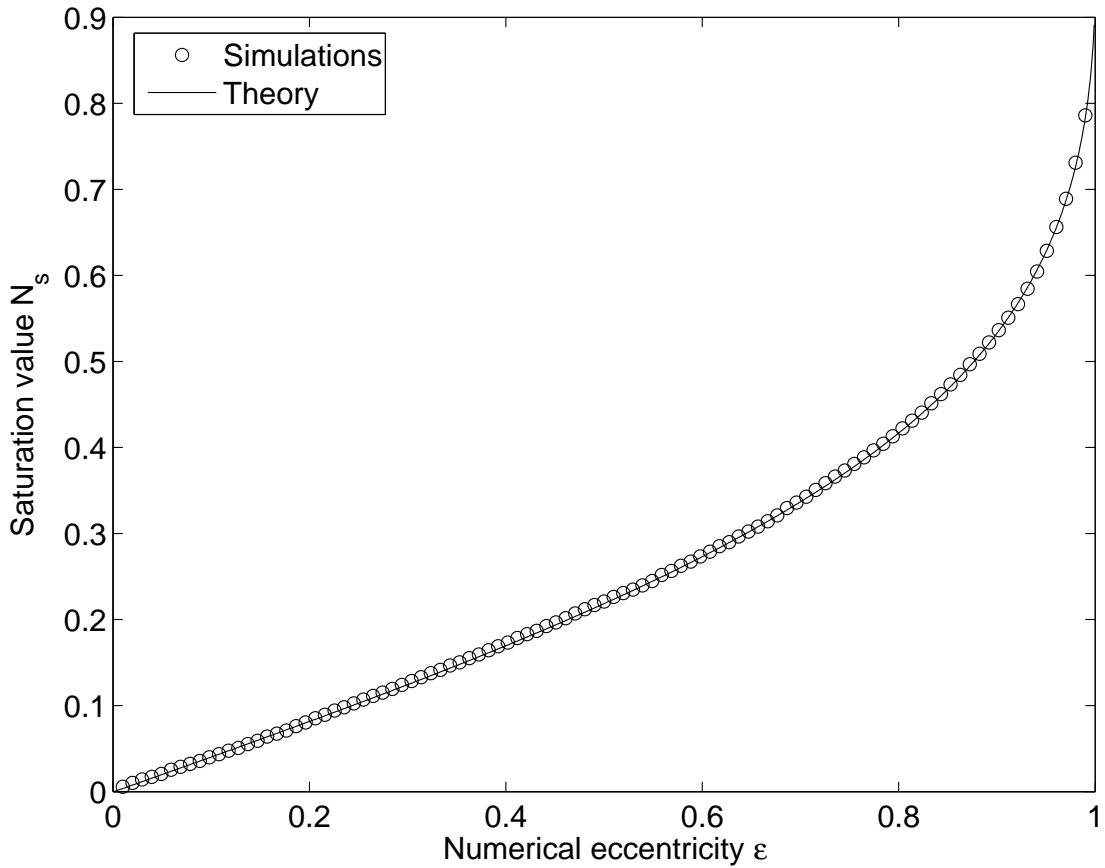}
        \caption{Dependence of the saturation value $N_s$ on
$\varepsilon$.}\label{fig:StTheoSim}
\end{figure}
The fraction of escaped particles as a function of $\varepsilon$ is shown in
Fig. \ref{fig:StTheoSim}. One can also calculate $N_s(\varepsilon)$ from
the following theoretical considerations, the result
is excellent agreement with the numerical data points, see Fig.
\ref{fig:StTheoSim}.

All initial conditions corresponding to particles propagating
on  librators lie inside the area $A_{II}(\varepsilon)$
bounded by the separatrix of Fig. \ref{fig:ElPSS}. We denote by
$f(\varepsilon) = A_{II}(\varepsilon)/A_{PSS}$ the ratio of
$A_{II}(\varepsilon)$ and the total area $A_{PSS}$ of the phase
space.
\begin{equation}\label{eq:ElAreaPSS}
    A_{PSS} = (p_{max}-p_{min})\cdot(\varphi_{max}-\varphi_{min}) =4\pi.
\end{equation}
To calculate $A_{II}(\varepsilon)$, we need an analytic expression
of the curve belonging to the upper half of the separatrix, that
is a function $p_{sx}(\varphi)$. Then, this area is given by
\begin{equation}\label{eq:ElAreaII}
    A_{II}(\varepsilon) = 2\int_{0}^{2\pi} d \varphi
    \,p_{sx}(\varphi).
\end{equation}
We know that for the
motion along the separatrix $F(\varphi, p) = 0$. Thus, we can
exploit  \eqref{eq:El2ndIntPhi} and get
\begin{equation}\label{eq:ElSeparatrix}
    p_{sx}(\varphi) = p(\varphi, F=0) = \sqrt{\frac{\varepsilon^2}{1+(1-\varepsilon^2)\cot^2
    \varphi}}.
\end{equation}
We see immediately that $A_{II}(\varepsilon)$ \eqref{eq:ElAreaII}
depends on $\varepsilon$, and so does the saturation value. To obtain
$N_s(\varepsilon)$, we have to account for the fact that the initial conditions
are distributed uniformly in the $\alpha,\,\varphi$ - and not the $p,\,\varphi$
- space. Hence,
\begin{equation}\label{eq:ElSepAlpha}
    \alpha_{sx}(\varphi) = \arccos (p_{sx}(\varphi)) = \arccos
\sqrt{\frac{\varepsilon^2}{1+(1-\varepsilon^2)\cot^2 \varphi}}
\end{equation}
and as a result $f'(\varepsilon)= A'_{II}(\varepsilon)/2\pi^2$,  similar to
\eqref{eq:ElAreaPSS} and $A'_{II}(\varepsilon)$ is
\begin{equation}\label{eq:ElAreaIIPrime}
    A'_{II}(\varepsilon) = 2\int_{0}^{2\pi} d \varphi
    \,\alpha_{sx}(\varphi).
\end{equation}
The fraction of escaped particles is just $1-f'$ and the
saturation number is
\begin{equation}\label{eq:ElEscSaturation}
    N_s(\varepsilon) = f'(\varepsilon).
\end{equation}
In Fig. \ref{fig:StTheoSim}, perfect agreement between the above
presented analytical considerations and the numerical simulations can be seen. As a
consequence, \textit{varying $\varepsilon$ allows us to control the
number  of particles being emitted}.

\section{Time-dependent Ellipse}\label{ch:TimeDepEllipse}

In this section, we investigate the escape rates for the time-dependent
ellipse. Since the boundary transfers momentum to the particles
upon collisions, their energy  is not conserved any
more. The collision point of a particle with the
boundary is not defined by $\varphi$ only, but we need additionally the
time $t$ to make the point well-defined in coordinate space, since
the boundary $\mathcal{B}(t)$ depends explicitly on $t$.
Likewise, the direction of a particle has to be described by
$\bm{v}=(v_x,v_y)$ and not just by $p=\cos\alpha$, since
$|\bm{v}|\neq \text{const}$. Representative visualizations of the resulting 4D-phase
space, like the 2D-PSS for the static billiard, are difficult to achieve, if not
to say prohibitive.

To drive the ellipse, we apply harmonic  oscillations to its
boundary $\mathcal{B}(t)$
\begin{equation}\label{eq:SetBoundary}
    \mathcal{B}(t)= \left\{\bm{b}(\varphi,t)|\,\varphi\in [0,2\pi)\right\}
\end{equation}
\begin{equation}\label{eq:EllBoundary}
 \bm{b}(\varphi,t)= \left( \begin{array}{c} x(\varphi,t)\\ y(\varphi,t)\\
\end{array}\right) = \left (\begin{array}{c} A(t) \cos \varphi \\
   B(t) \sin \varphi\\ \end{array}\right )
\end{equation}
where $A(t)$ and $B(t)$ are given by
\begin{subequations}\label{eq:Driving}
\begin{eqnarray}
  A(t) &=& A_0 + C \sin (\omega t+ \delta) \\
  B(t) &=& B_0 + C \sin (\omega t+ \delta)
\end{eqnarray}
\end{subequations}
$C>0$ is the driving amplitude and $\delta$ is a phase shift.
$A_0, B_0$ and $C$ have to be chosen in a way, that $A(t)>0$ and
$B(t)>0$ for all $t$. We refer to \eqref{eq:Driving} as the
\textit{breathing ellipse}. As already done in section
\ref{ch:StaticEllipse}, we set $A_0=2$ and $B_0=1$, and use values of $C$ between $0.01$ and
$0.30$ only. The velocity $\bm{u}(\varphi,t)$ of the boundary and the numerical eccentricity are
\begin{equation}\label{eq:BoundaryVelocity}
    \mathbf{u}(\varphi,t)  = \left(\begin{array}{c}
      \omega C \cos (\omega t + \delta) \cos \varphi \\
      \omega C \cos (\omega t + \delta) \sin \varphi \\
    \end{array}  \right),
\end{equation}
\begin{equation}
\varepsilon (t) =  \sqrt{1-\frac{(1 + C \sin (\omega t+ \delta))^2}{(2 + C \sin (\omega t+ \delta))^2}}.
\end{equation}

\subsection{Mapping}\label{ch:TDMapping}
Just like in the static case, a discrete
mapping is sufficient to characterize  the full dynamics of a
particle.  Consequently, the trajectory of a particle
consisting of $N$ bounces is given by
\begin{equation}\label{eq:TDSet}
    \mathcal{C} = \{(t_0, \varphi_0,\bm{v}_0), (t_1, \varphi_1,\bm{v}_1), \dots,
(t_N, \varphi_N,\bm{v}_N)\}.
\end{equation}
The mapping for the next collision time $t_{n+1}$ is determined
implicitly by
\begin{equation}\label{eq:Implicit2}
    \left ( \frac{v_x^n (t_{n+1}-t_n) + x_n}{A_0 + C \sin (\omega t_{n+1}+
\delta)}
\right )^2 + \left (\frac{v_y^n (t_{n+1}-t_n) + y_n}{B_0 + C \sin (\omega
t_{n+1}+
\delta)} \right )^2 -1 = 0
\end{equation}
where for a given $t_n$ and $\varphi_n$, $x_n$ and $y_n$ are
calculated from \eqref{eq:EllBoundary} and $t_{n+1}$ is defined by
the smallest $t_{n+1}>t_n$ that solves \eqref{eq:Implicit2}. The
next collision point is given by $\bm{x}_{n+1}=\bm{x}_n+\bm{v}_n
(t_{n+1}-t_n)$ and $\varphi_{n+1}$ can be obtained by inverting
\eqref{eq:EllBoundary}. Once
$(t_{n+1},\varphi_{n+1})$ is determined, the next velocity $\bm{v}_{n+1}$ is
given by
\begin{equation}\label{eq:NextVelocityTime1}
    \bm{v}_{n+1} = \bm{v}_n -2 \left [ \hat{ \bm{n}}_{n+1} \cdot (\bm{v}_n -
\bm{u}_{n+1})\right ] \cdot \hat{ \bm{n}}_{n+1}
\end{equation}
where the boundary velocity $\bm{u}_{n+1}$ is given by
\eqref{eq:BoundaryVelocity} and the  the normal vector by
$\hat{\bm{n}}_{n+1}=\hat{\bm{n}}'_{n+1}/|\hat{\bm{n}}'_{n+1}|$,\,
$\hat{\bm{n}}'_{n+1}=(-B(t_{n+1}) \cos \varphi_{n+1},-A(t_{n+1})
\sin \varphi_{n+1})^\top$.

The maximal  velocity change of a particle upon a single collision
with the boundary is according to eq. \eqref{eq:NextVelocityTime1}
$\triangle |\bm{v}| = \pm 2\omega C$. Thus, it can happen that the
particle undergoing a boundary collision at the time $t'$ is not
reflected, in a sense that the sign of the velocity component
normal to the boundaries tangent is not reversed, but continues
traveling outside $\mathcal{B}(t')$, but of course still inside
$\mathcal{B}(t>t')$. This is the case if the ellipse is expanding
and $u_n < v_n < 2u_n$ holds, where $v_n$ and $u_n$ are the normal
component of the particle and the boundary velocity before the
collision. As a consequence, the angle
$\alpha$ between the tangent $\bm{t}$ and the velocity $\bm{v}$ is
not restricted to the interval $[0,\,\pi]$ as it was  in the case of the
static ellipse, but now $\alpha \in [-\pi,\,\pi]$. Upon such  collisions
with the expanding boundary, the particles are always slowed down,
they lose energy  \cite{Pa02,Pa04}, whereas upon collisions with the contracting
ellipse they gain energy.

\subsection{Escape Rates}\label{ch:TDEscapeRate}
We focus on the case $s_\triangle = 0$, in order to  examine the
effect of the driving on the number of particles in the billiard.
For the static case, the saturation value was caused by the librator orbits,
see section \ref{ch:SatValue}. We assume that these librator
orbits will be deformed or partially destroyed by the driving,
leading to a non-vanishing decay even for large times. On the
other hand, we expect no stabilization of the rotator orbits, i.e.
no deformation in a way that they will not escape. All periodic
orbits become unstable when applying the driving and in ref.
\cite{Koi95b} it was concluded, that it is impossible to trap
unstable periodic orbits in the ellipse via boundary oscillations. Note that this is not true in general for driven systems, unstable periodic orbits can be stabilized by a driving force, e.g. in the Kapitsa pendulum  \cite{Ka51}.

\begin{figure}[ht]
        \includegraphics[width=0.9\columnwidth]{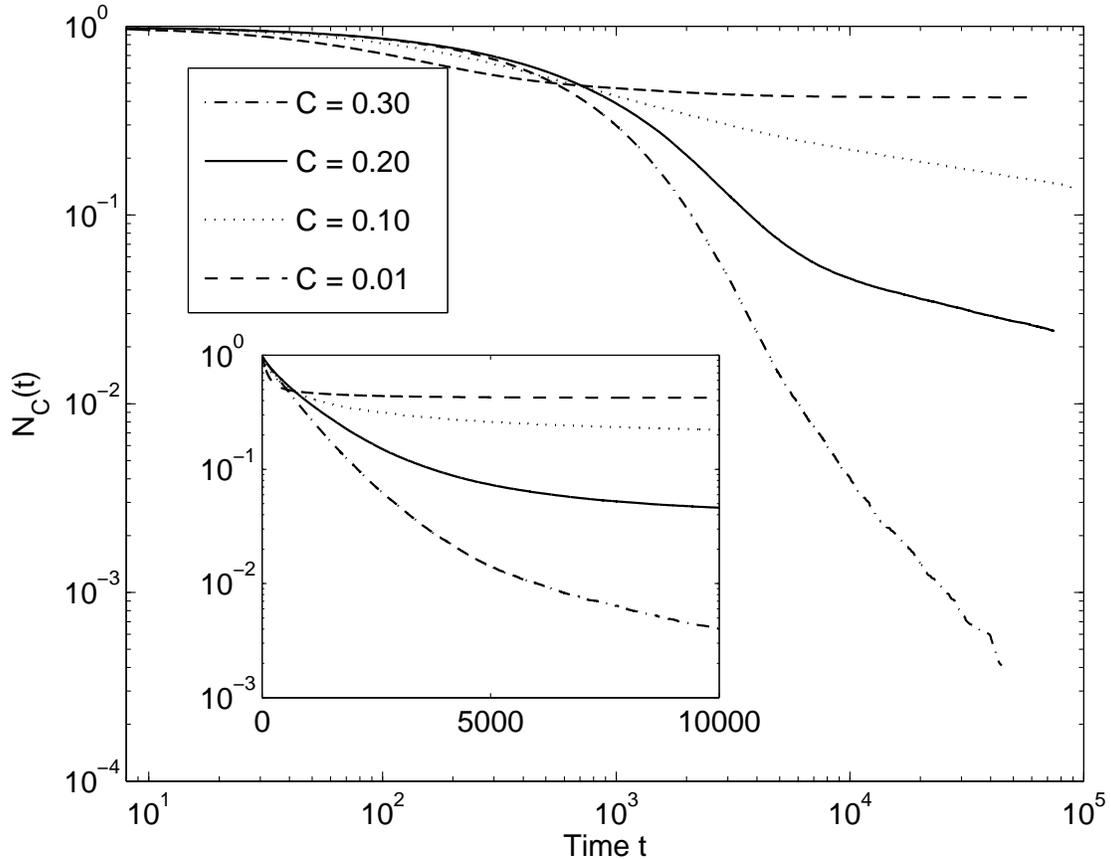}
        \caption{Fraction of remaining particles in the IVE as a function of
time for different values of the amplitude $C$, semi-logarithmic plot in the
inset.}\label{fig:EscSL_IV}
\end{figure}

\begin{figure}[ht]
        \includegraphics[width=0.9\columnwidth]{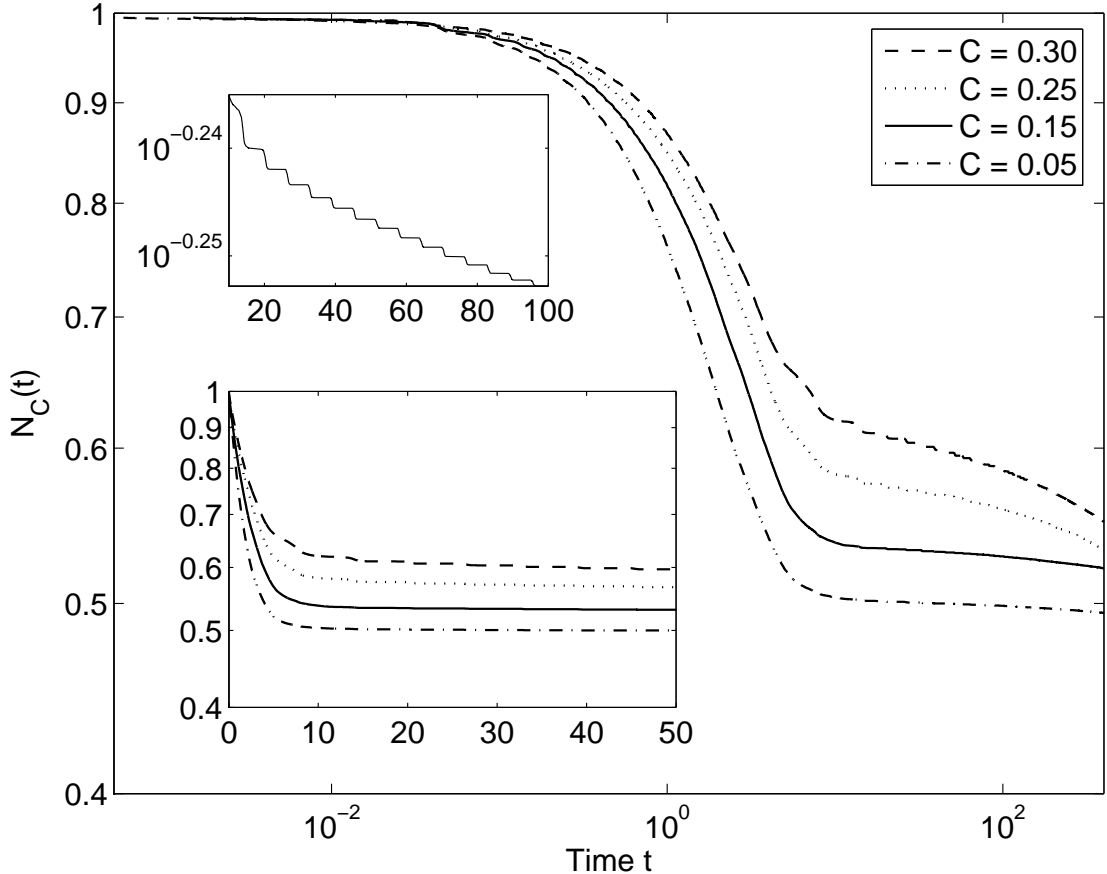}
        \caption{Same as Fig. \ref{fig:EscSL_IV} for the HVE, additionally
$2\pi$-oscillations of the decay are shown in the inset.}\label{fig:EscSL_HV}
\end{figure}
We consider two different borderline cases: $|\bm{v_0}|\approx
\omega C$ (intermediate velocity ensemble (IVE)) and
$|\bm{v_0}|\gg \omega C$ (high velocity ensemble (HVE)). In the
first case, the velocity has the same order of magnitude as the
the boundary velocity. This leads to
a momentum transfer (maximal $2\omega C$) of the same order of
magnitude compared to the initial momentum  and we expect significant changes in
the dynamics. In
the second case, the particles move much faster than the boundary,
consequently the momentum transfer will be very small, the
dynamics will be similar to the one of section
\ref{ch:StaticEllipse}.  Naturally, it would be also interesting
to examine the case $|\bm{v}_0|\ll \omega C$. However, the first
few collisions then accelerate the particles to velocities
$|\bm{v}|\approx \omega C$ and, after a short time, we encounter
the situation of the first case. The parameters of the simulations
are $N_0 = 10^5$, $\omega =1$,
$C=0.01,0.05,0.10,0.15,0.20,0.25,0.30$ and $\delta =0$. To ensure
that all the particles move inside the billiard, we let them start
on the smallest ellipse (for a given $C$), the initial position $\varphi_0$ and the
initial angle $\alpha_0$ are chosen randomly. The initial velocity
$\bm{v}_0$ is given by $\bm{v}_0 = (\cos \alpha_0,\sin \alpha_0)$
(IVE) and $\bm{v}_0 = 100\cdot(\cos \alpha_0,\sin \alpha_0)$ (HVE)
respectively. The fraction of remaining particles $N_C(t)$ as a
function of time  for different amplitudes $C$ is shown in Figs.
\ref{fig:EscSL_IV} (IVE) and \ref{fig:EscSL_HV} (HVE).

Firstly, we describe the behavior of the IVE. We observe a short
but fast decay ($t< 500$), followed by a transient ($500<t<5000$)
in which the decay slows down and for $t>5000$, the decay is much
slower than the initial fast decay for all  values of $C$. At $t=
10^4$, the values of the fraction of remaining particles are
ordered according to the driving amplitudes, the lower $C$ is, the
higher is $N_C(t=10^4)$, i.e. $N_C(t=10^4)$ depends monotonically
on the amplitude $C$.  For $t>10^4$, $N_C(t)$ does not stay
constant, but is still decreasing. The absolute value of the
emission rate $\dot{N}_C(t)$ is the larger the larger $C$ is. This
can be seen nicely  in the double-logarithmic plot of Fig.
\ref{fig:EscSL_IV}. For values of $t$ between $10^4$ and $10^5$,
we encounter approximately an algebraic decay $N_C(t) \sim t^{-w}$
(we remark that this algebraic decay has been numerically shown to
exist for much longer times than illustrated in Fig.
\ref{fig:EscSL_IV}), where the decay constant $w$ increases
monotonically with increasing $C$ (this fact is based not only on
the four values of the driving amplitude $C$ shown here, but on
simulations carried out for 20 values of $C$ between $0.01$ and
$0.30$).

The subdivision of the behavior, into fast initial decay -
transition period - slow (near algebraic) decay, is even more
pronounced in the case of the HVE, see the inset of Fig.
\ref{fig:EscSL_HV}. An exponential decay for small values of $t$
slows down at around $t\approx 5$  and the fraction of remaining
particles  seems to approach a constant value. From Fig.
\ref{fig:EscSL_HV} however we see, that the fraction of remaining
particles    decays for $t>10$ roughly according to an algebraic
decay (at least for small values of the driving amplitude) $N(t)
\sim t^w$ with a decay constant $w$. If we compare the fraction of
remaining particles at $t=50$ for different values of the
amplitude (inset of Fig. \ref{fig:EscSL_HV}), we see that they are
monotonically ordered according to the driving amplitudes.
Surprisingly,   most of the particles remain within the billiard
in case of the largest driving amplitude $C=0.30$ and the smallest
fraction remains in case of the smallest amplitude $C=0.05$.  The
explanation of this effect is provided later, in section
\ref{ch:AngMom}, when we examine the dependence of the PAM
$F(\varphi,p)$ on the driving.

In the inset of Fig. \ref{fig:EscSL_HV}, a modulation of the
escape rate with period $T=2\pi$ can be seen, being exactly the
period of the applied driving law \eqref{eq:EllBoundary}.
Specifically, for $t \ge 10$, where all particles starting on
rotator orbits have already escaped, $N_C(t) \approx const. $
during approximately $11/12$ (empirically observed) of one period
and subsequently $\dot{N}_C(t) \neq 0$ during a time interval
$T/12$ only. From this behavior it is evident that the ellipse
operates from a certain time on as a pulsed source of particles.
These repeated intervals are centered around points $t_m$ of
maximal extension of the ellipse, $t_m= (4m +1)\pi/2,\, m=
2,3,4,\dots$ During the expansion period, dominantly vertical but
also horizontal processes turn librators into rotators. The moving
ellipse remains for a comparatively long time period in the
vicinity of the extremal configuration at $t_m$ and consequently
the newly created rotators escape. Therefore, the dynamics is
effectively probed during these short time intervals centered
around $t_m$. During the contraction period, the librators are
stabilized via  vertical processes, consequently $\dot{N}_C(t)
\approx 0$ during $11/12$ of a period $T$.

\subsection{Mechanisms for the destruction of the
Librators}\label{ch:DesTypII}
In the driven ellipse librators can escape from
the billiard whereas this is not the case for the static ellipse.
There are two fundamental processes that perturb or even
completely destroy the librator orbits (unprimed variables denote
the static, whereas primed ones describe the driven system):
\begin{enumerate}
    \item \textit{Vertical process:} The angle of incidence of a collision does
not coincide with the reflection angle because of a change of momentum due to
the motion of the boundary of the ellipse. In phase space, the momentum then
undergoes a certain change $\triangle p$ upon a collision and the particle moves
vertically in the PSS.
    \item \textit{Horizontal process:} A particle that would hit the boundary at
a certain point $\varphi$, hits the boundary in the driven case at $\varphi'$,
simply because the ellipses boundary has moved, whereas $p$ stays nearly
unchanged. This corresponds to a horizontal move in the PSS.
\end{enumerate}
These processes are fundamental in the sense that every change
$\triangle F$ can be decomposed (at least for small changes $(\triangle
\varphi,\, \triangle p)$) into $\triangle F = \triangle
F_h+\triangle F_v$, where $\triangle F_{h,v}$ denote the
individual changes caused by the horizontal, the vertical process,
respectively.

 In general, these effects do not appear isolated but
a combination $(\triangle \varphi,\, \triangle p)$ of both will
occur in a single collision. We can compare the orbits
$(\varphi_i',p_i')$ of the driven ellipse to the corresponding
ones of the static ellipse $(\varphi_i,p_i)$ by considering the
quantity $F(\varphi,p)$ (see eq. \eqref{eq:El2ndIntPhi}). In
contrast to the case of the static ellipse where
$F(\varphi_i,p_i)=const. \,\forall \,i$ we have
$F(\varphi_i',p_i') \neq F(\varphi_j',p_j') \, (i\neq j)$ for the
driven case, i.e. $F$ is no longer a constant of motion. The
difference $\triangle F$ (see Fig. \ref{fig:ElPSS}) upon a
collision is a measure of whether a librator approaches  the
separatrix ($\triangle F > 0$) or whether it moves in phase space
towards the position of the elliptic fixed points ($\triangle F <
0$) of the static case. An increase with respect to $F$ reflects
the dependency of 'moving' in phase space from confined librator
to escaping rotator orbits.

\subsubsection{Vertical Processes}\label{ch:VerticalProc}
\begin{figure}[ht]
        \includegraphics[width=0.9\columnwidth]{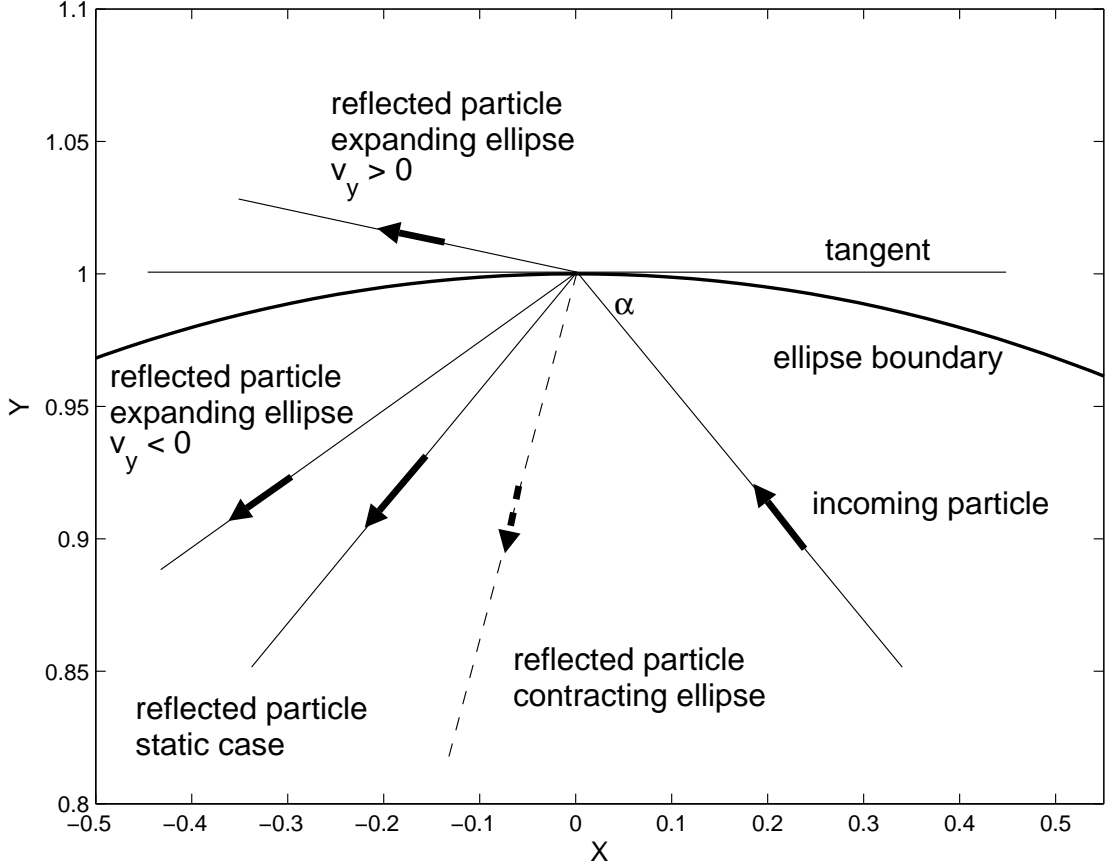}
        \caption{Vertical process in coordinate
space.}\label{fig:Vertical_process_Coord}
\end{figure}
\begin{figure}[ht]
        \includegraphics[width=0.9\columnwidth]{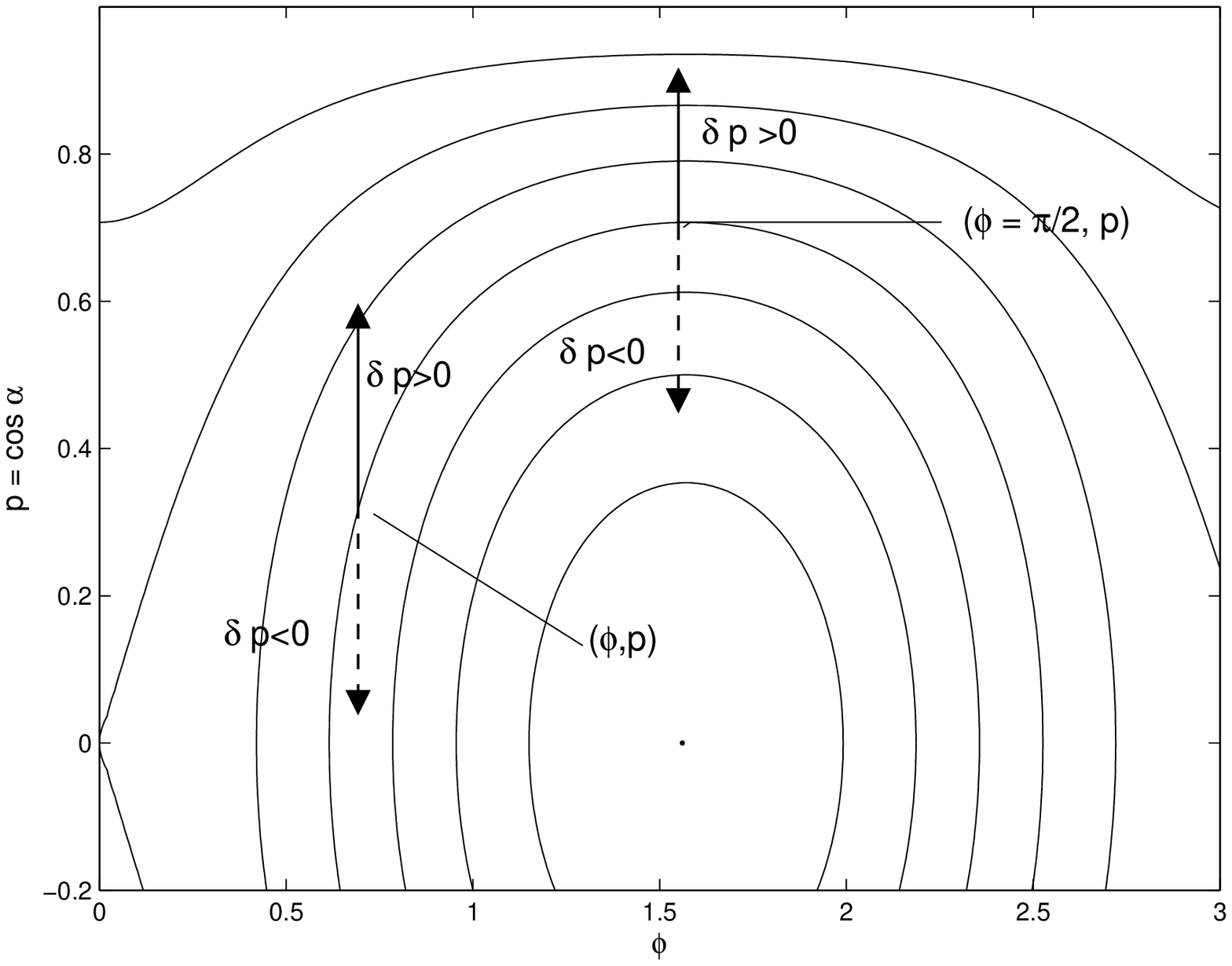}
        \caption{Vertical process in phase space.}\label{fig:Vertical_process_PhaseSpace}
\end{figure}
To isolate this effect, we examine a particle that hits the
boundary at $\varphi=\pi/2$ under a certain angle $\alpha$ in the
static case. The velocity of the particle can be written as
$\bm{v}_0= (-v \cos \alpha, v \sin \alpha), \,v=|\bm{v}_0|$. In
the driven ellipse, we will assume that the particle hits the
boundary in the neutral position ($A(t)=A_0,\,B(t)=B_0$), so we
have $\varphi' = \varphi = \pi/2, \Rightarrow \triangle \varphi
=0$.  The boundary velocity $\bm{u}(\varphi,t)$ of the ellipse is
maximal at this configuration and has a vertical component only,
$u_n=u_y = \pm \omega C$, depending on whether the ellipse is
expanding ``+'' or contracting ``-''. The of the
particle at the next collision in the static case is $ \bm{v}_1 =
(-\cos \alpha, -\sin \alpha )^\top$ and hence  $p_1 = p_0 = \cos
\alpha$.  The corresponding velocity in the driven case is $\bm{v}'_1 = (-v
\cos \alpha, -v\sin \alpha \pm 2 \omega C)^\top$. Now, $p_0'=p_0
\neq p_1'$, since
\begin{equation}\label{eq:CosAlphaNew}
   p'_1 = \frac{\cos \alpha}{\underbrace{\sqrt{1\mp \frac{4\omega C}{v} \sin
\alpha + \frac{4\omega^2 C^2}{v^2}}}_{f}}
\end{equation}
If the ellipse is contracting (``+'' sign in the factor $f$),
$p'_1= \cos \alpha /f$ is smaller than $p_1=\cos \alpha$ because
$f>1$, i.e. $\triangle p <0$ ($p'_1 = p_1 + \triangle p$). In
phase space the particles `moves' towards the elliptic fixed
points, which does not lead to the destruction of the librators. If the
ellipse is expanding (``-'' sign in the factor
$f$), $\triangle p $ will be larger than zero if $f<1$. This is
equivalent to
\begin{equation}\label{eq:fsmaller1}
    \sin \alpha > \frac{\omega C}{v}.
\end{equation}
and since $v_n = v \sin \alpha$ and $u_n = \omega C$ eq.
\eqref{eq:fsmaller1} is equivalent to $v_n > u_n$, which is a
necessary condition for a collision to take place. If additionally
to eq. \eqref{eq:fsmaller1} $\sin \alpha < 2\omega C/v$, then $v_n
= v_y >0$ and thus $\alpha'<0$, see Fig.
\ref{fig:Vertical_process_Coord}, the particle is not
reflected in the sense that the sign of $v_n$ is not reversed. For
$\sin \alpha > 2\omega C/v$ it follows that $v_n=v_y <0$, the
particle is reflected. In both cases $\triangle p>0$,  the
particle moves towards or even beyond the separatrix; in the
latter case, the librator has changed into a rotator.
This process can therefore lead to the destruction of the librator
orbits. In Figs. \ref{fig:Vertical_process_Coord} and
\ref{fig:Vertical_process_PhaseSpace}, the process is shown in
coordinate, as well as in phase space. This process happens  for
any value of $\varphi$, nevertheless, the case $\varphi \approx
\pi/2$ is for two reasons especially important:
\begin{enumerate}
    \item The absolute value of the change $\triangle p$ is largest for $\cos
\alpha \approx \sin \alpha$, for particles on librators this is
approximately true if $\varphi \approx \pi/2$.
    \item For a constant  $\triangle p$, the corresponding $\triangle
F(\varphi,p)$ is largest for $\varphi=\pi/2$, because the vertical spacing of
the invariant curves is smallest at $\varphi =\pi/2$.
\end{enumerate}
Consequently, the horizontal processes that contribute the most to
changing a librator into a rotator occur mainly around
$\varphi \approx \pi/2$ (and  $\varphi \approx 3\pi/2$
because of symmetry).

\subsubsection{Horizontal Processes}\label{ch:HorizontalProc}
\begin{figure}[ht]
        \includegraphics[width=0.9\columnwidth]{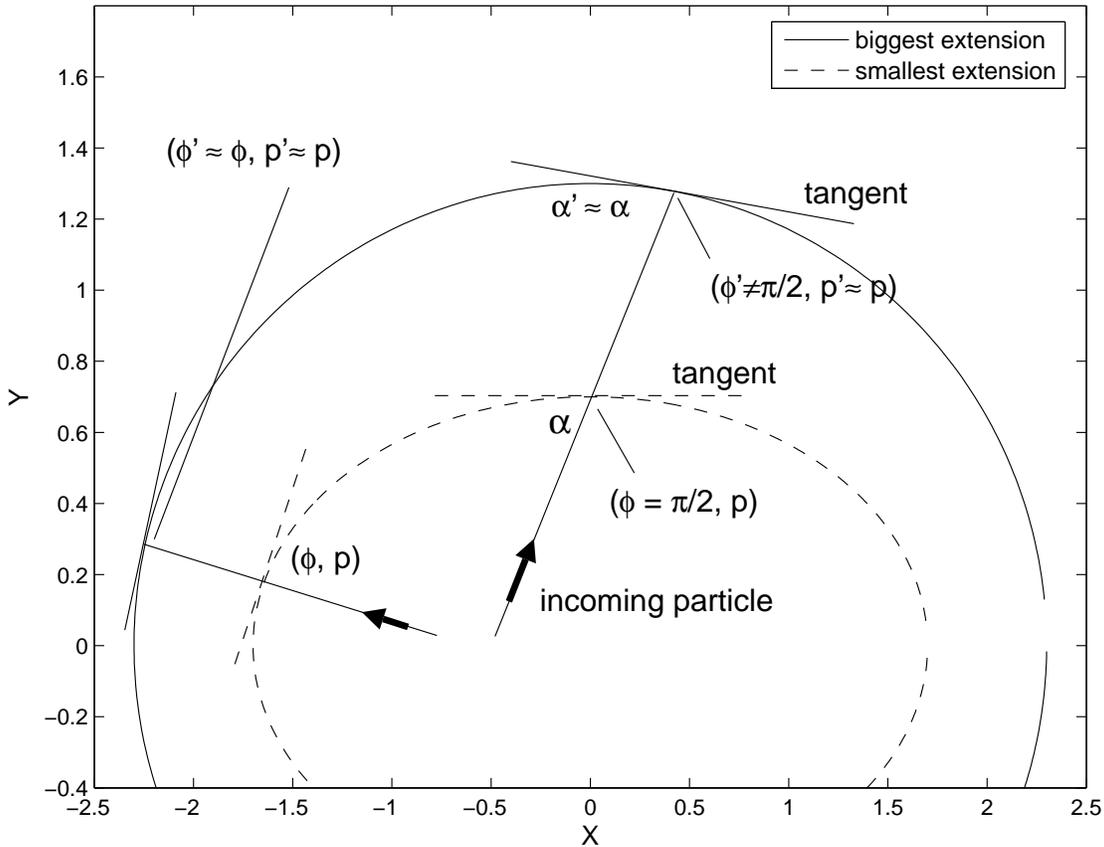}
        \caption{Horizontal process in coordinate space.}\label{fig:Horizontal_process_Coord}
\end{figure}
\begin{figure}[ht]
        \includegraphics[width=0.9\columnwidth]{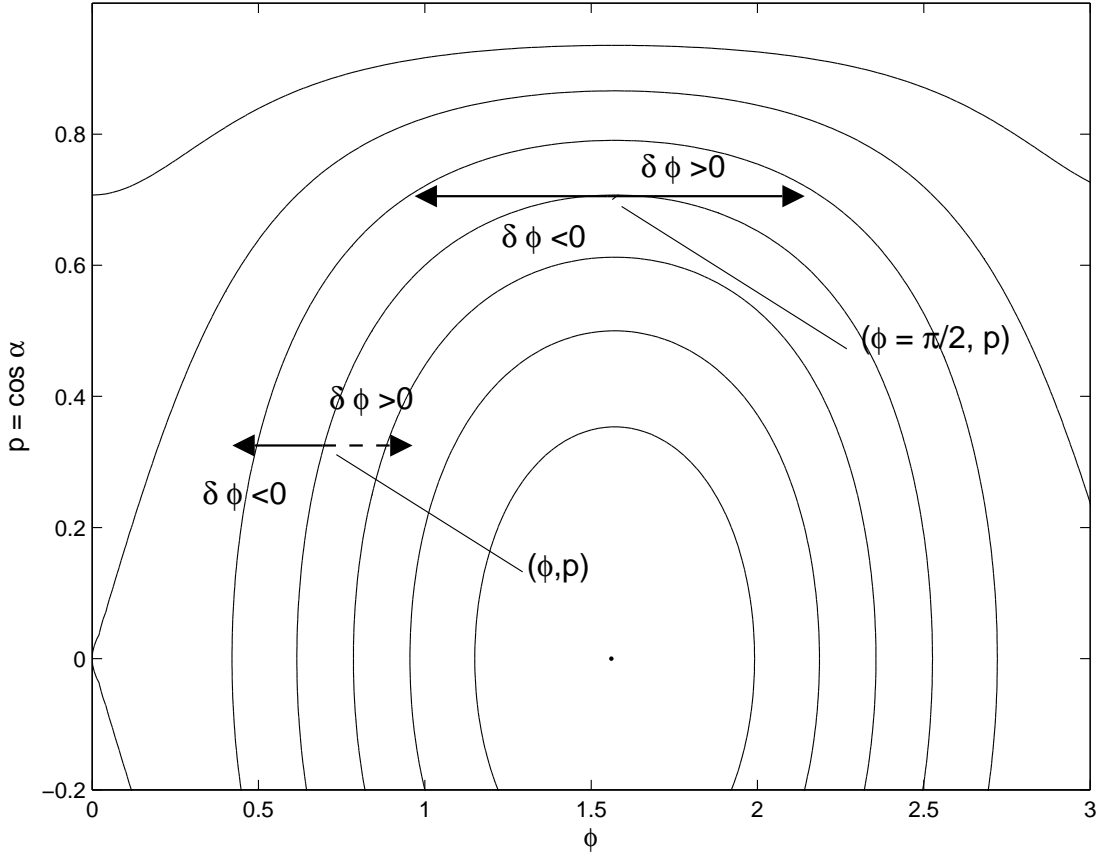}
        \caption{Horizontal process in phase space.}\label{fig:Horizontal_process_PhaseSpace}
\end{figure}
We consider a particle that starts moving in a static ellipse
corresponding to the smallest possible ellipse (given a certain amplitude) in
the driven case, hitting
the boundary of this static ellipse at $\varphi = \pi/2$ under a
certain angle $\alpha$, see Fig.
\ref{fig:Horizontal_process_Coord}. We choose all the parameters such that
the particle will hit the boundary of the driven
ellipse at the time $t'$ at the position $\varphi'\neq \varphi$,
when the boundary $\mathcal{B}(t')$ has its maximal extension. The new
angle $\alpha'$ is approximately equal to $\alpha$, see Fig.
\ref{fig:Horizontal_process_Coord}. This rough estimate  becomes
better  with increasing $\alpha$ and decreasing distance between
the two ellipses, thus  $\triangle p \approx 0$, i.e. the particle moves
horizontally in the PSS, see Fig. \ref{fig:Horizontal_process_PhaseSpace}. To
calculate $\varphi'$ or $\triangle \varphi$ as a function of $C$ and
$\alpha$ is very tedious and the exact result is not very helpful.
We therefore restrict  ourselves to an approximation and linearize the
boundary of the ellipse locally at $\varphi=\pi/2$. The collision point
on $\mathcal{B}(t)$ is
\begin{equation}\label{eq:PhiDelta}
    \varphi' = \arctan \frac{y}{x} \approx \arctan \frac{(2+C)\cdot\tan \alpha }{2C}.
\end{equation}
In general we have to respect the sign of $x$ and $y$ to obtain
$\varphi'$ from \eqref{eq:PhiDelta}. $|\triangle \varphi| =|
\varphi' - \varphi|$ depends sensitively on $\alpha$ and
decreases with increasing $\alpha$. $|\triangle \varphi|$ is
largest for $\varphi \approx \pi/2$, since  $\alpha$ there
reaches its minimal value for the librator orbits. The sign of  the
corresponding change $\triangle F$ depends on the sign of
$\triangle \varphi$ and the quadrant in which $\varphi$ lies (e.g.
for $\triangle \varphi <0$ $\triangle F >0$ if $\varphi$ lies in
the second or forth quadrant). There is no obvious region with respect to $\varphi$
where the destruction of the librators occurs mainly, since two
opposite effects balance each other:
\begin{enumerate}
    \item The absolute value of the change $|\triangle \varphi|$ is largest for
$\varphi \approx \pi/2$.
    \item Given a certain $\triangle \varphi$, the corresponding changes
$\triangle F$ increase with increasing distance between $\varphi$ and $\pi/2$,
because the horizontal
spacings of the invariant curves is smallest there, see Fig.
\ref{fig:Horizontal_process_PhaseSpace}.
\end{enumerate}
Nevertheless, the position $\varphi = \pi/2$ (and $\varphi =
3\pi/2$) is in a way exceptional: any change $\triangle\varphi$,
independent of the sign, results in a change $\triangle F >0$,
leading to a destruction of the librators.

\subsection{Qualitative Model of the Decay}\label{ch:model}
We isolated two processes that are able to change $F(\varphi,p)$ in the course
of the dynamics. Particles on librator orbits are scattered upon boundary
collisions either towards the elliptic fixed points or towards or even beyond
the
separatrix.  For a single particle,
such a scattering  process  happens at every collision, the
effective change $\triangle F$ after a certain time depends on the
sequence of these processes, hence $\triangle F = \triangle F_1 +
\triangle F_2+ \dots +\triangle F_n$ after $n$ collisions. This
effective change in $F(\varphi,p)$ is very difficult, if not
impossible, to predict, since each individual change $\triangle
F_i$ depends on four parameters already: 1. the absolute value
$|\bm{v}_i|$ of the particle velocity, 2. the angle $\alpha_i$ of
the velocity with the boundary, 3. the location $\varphi_i$ of the
collision point on the boundary and 4. the time $t_i$ which
determines the position of the boundary and the boundary velocity
$\bm{u}_i(t_1)$ of the ellipse (these four parameters are of course just
the variables of the four-dimensional discrete mapping, see section
\ref{ch:TDMapping}). Now we consider not only a single particle, but an ensemble
of $N$ particles with initial conditions $(\varphi_j,\,\bm{v}_j), \,j=1,2,\dots
N$. The effective change $(\triangle F)_j$ (where the index $j$ indicates the
$j$th particle) after $n$ collisions can vary significantly from particle to
particle, since the sequence of these four parameters will be very differently
for each individual particles. Each of the $N$ sequences is governed  by
applying the discrete mapping of section \ref{ch:TDMapping} $n$-times on
each initial condition $(\varphi_j,\,\bm{v}_j)$. The underlying
nonlinear dynamics of this discrete mapping and the fact that all particles
start from different initial conditions leads to such  unique sequences and
will cause consequently large fluctuations in the effective $\triangle F$ and
accordingly large fluctuations in quantities that  depend on $F(\varphi,p)$. In
the following, a qualitative explanation of the escape rates $N_C(t)$
(decay) of the IVE and the HVE is given.

We focus on the HVE first. The initial fast decay of the number of particles
($t<5$) is due to the rotator orbits that are connected with the hole and
escape
very rapidly. Additionally some of the particles starting on librator orbits
near
the separatrix $F\lessapprox0$ contribute. The longer-time decay ($t>10$) is
caused by particles starting on librators that have been scattered
across
the separatrix.  The closer an orbit of a particle lies near the
elliptic
fixed points, the longer it takes until the
effective change $\triangle F$ is big enough to reach the separatrix ($F=0$).
From equations \eqref{eq:CosAlphaNew} and \eqref{eq:PhiDelta} it
follows that the individual changes $\triangle F_i$ under  a single
collision increase with increasing amplitude $C$. This explains
the increasing emission rate $\dot{N}_C(t)$ with increasing $C$, since at a
given time $t$, the number of particles that can participate in the decay is
larger for larger values of $C$. The decay in  the transient region ($5<t<10$)
is caused by a superposition of the tail of the initial fast decay (roughly
exponential) and the onset of the slow (roughly algebraic) decay.

With very similar arguments, the decay of the IVE can be explained
qualitatively. Since the velocity of the particles and the velocity of the
boundary are of the
same order of magnitude, the changes $\triangle F$ are much larger compared to
the ones of the HVE. This leads to a very early onset of the slow (algebraic)
decay, consequently the transient region is broadened.

\section{Angular Momentum}\label{ch:AngMom}
To validate the qualitative model of section \ref{ch:model}, we
investigate the PAM $F(\varphi,p)$ further. The
contours of $F(\varphi,p)$ are shown in Fig. \ref{fig:ElPSS},
depending on the initial value of $F$, particles move on rotator or
librator orbits, see section \ref{ch:FundamentalProp}.

Through out the following sections, we  analyze properties such
as the escape time for an ensemble of particles possessing certain initial
distributions in e.g.
phase space or the PAM $F$. We emphasize that in case of the distribution of the $F$-values we
always refer to the initial distributions of $F$ at $t=0$
\footnote{\label{ftn1}Since the
initial conditions chosen in section \ref{ch:TDEscapeRate} do not lie on the
boundary of the ellipse, the first collision point with
the
boundary is taken, at this point $t\neq0$.}.

\subsection{Escape Time versus Initial Conditions in Phase Space}\label{ch:EscPSS}
\begin{figure}[ht]
        \includegraphics[width=0.9\columnwidth]{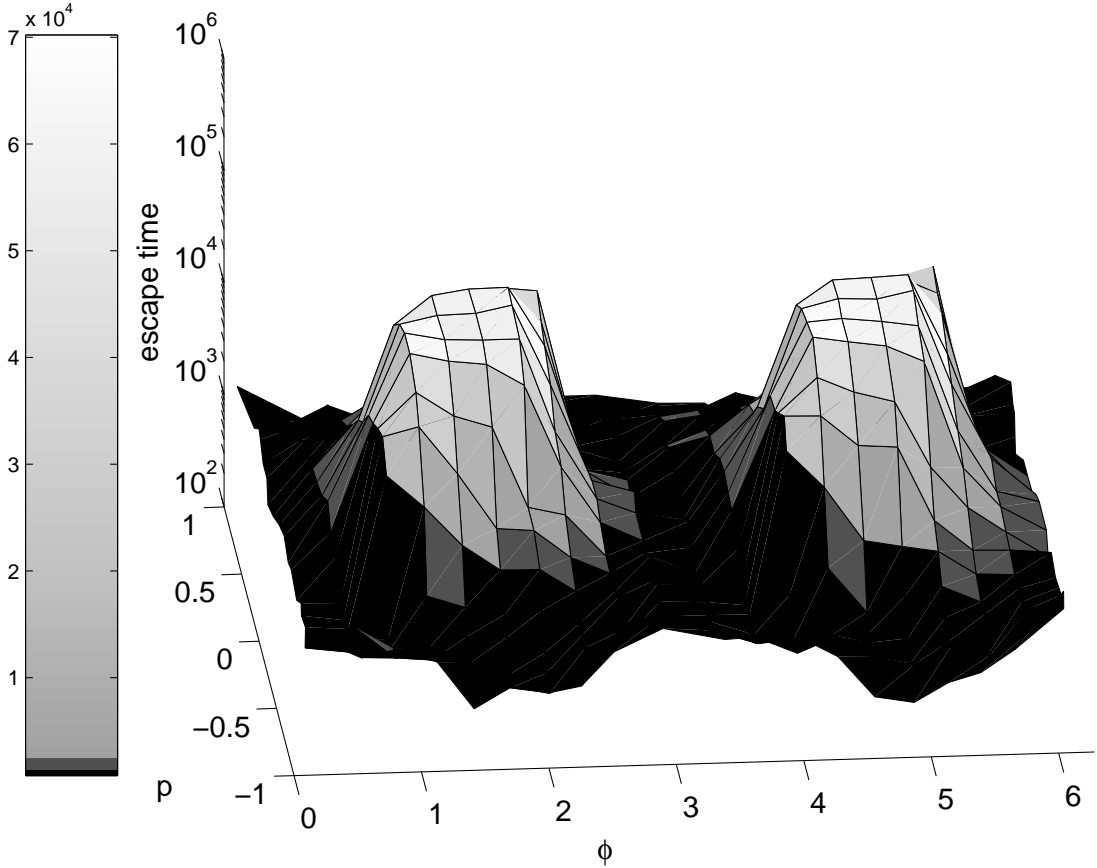}
        \caption{Escape time versus initial conditions in phase space for
$C=0.10$ (IVE). The escape time is largest for initial conditions close to the
elliptic fixed points.}\label{fig:EscTime_vs_PSS}
\end{figure}
Exemplarily, the escape time as a function of the starting points in phase space
is shown in Fig. \ref{fig:EscTime_vs_PSS} ($C=0.10$, IVE), i.e.
we assign to each initial condition $(\varphi_0,\,p_0)$ ($10^5$ particles) an
escape time.  Large values of the escape time correspond
to initial conditions belonging to librator orbits lying around the
elliptic fixed points at $(\varphi=\pi/2,\,p=0)$ and
$(\varphi=3\pi/2,\,p=0)$. On the other hand, initial conditions
corresponding to small values of the escape time lie around areas
which correspond to the rotator orbits.  Overall, the results of Fig.
\ref{fig:EscTime_vs_PSS} are in good agreement with our predictions of section
\ref{ch:model}, where we derived large escape times for
particles with initial conditions close to the two elliptic fixed
points and short escape times for particles starting on librator
orbits.

\subsection{Escape Time versus  Initial Angular Momentum}\label{ch:EscFCor}
In this section, we investigate the escape time $t_{esc}$ versus the
initial angular momentum $F(\varphi_0,p_0)$ of the corresponding
ensemble of particles in phase space for different amplitudes $C$. We consider
escaped particles only.
\begin{figure}[ht]
        \includegraphics[width=0.9\columnwidth]{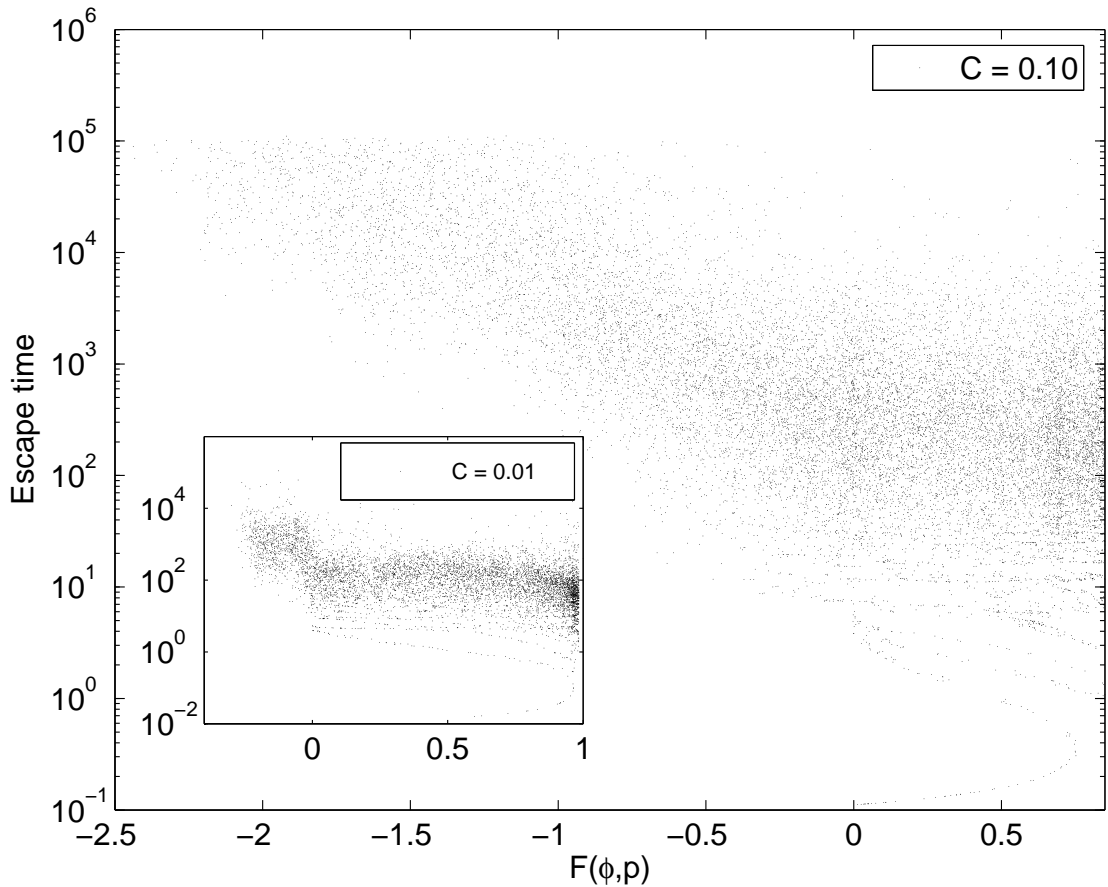}
        \caption{Escape time versus initial angular momentum $F(\varphi_0,p_0)$
($F=0$ corresponds to the separatrix in the static ellipse and $F=-3.1$ to the
elliptic fixed points) in the IVE for $C=0.10$ and
$C=0.01$ (inset).}\label{fig:EscTime_vs_F_Amp_001_IV}
\end{figure}
\begin{figure}[ht]
        \includegraphics[width=0.9\columnwidth]{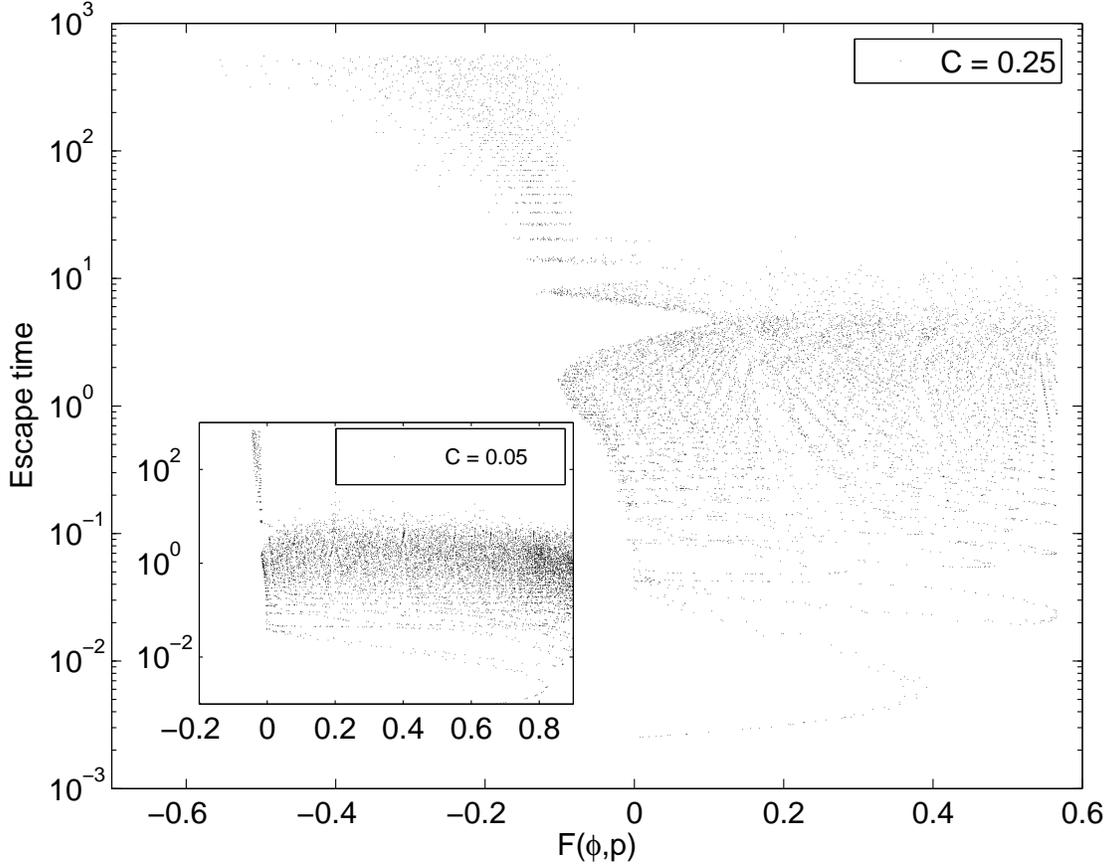}
        \caption{Same as Fig. \ref{fig:EscTime_vs_V_Amp_001_IV} for the HVE,
$C=0.25$ and $C=0.05$ (inset). }\label{fig:EscTime_vs_F_Amp_005_HV}
\end{figure}
The results of the IVE are shown in Fig.
\ref{fig:EscTime_vs_F_Amp_001_IV}. For small escape times $t_{esc}
\lesssim 10$, there is a narrow, serpentine chain in which all the
pairs $(t_{esc},\,F)$ lie. Only pairs with $F>0$ corresponding to
rotators occur for $t_{esc} \lesssim 10$. We will explain
this below in the course of the discussion of the HVE. Apart from this
narrow, serpentine chain, the values of the escape times for $F>0$
are scattered mainly over a rectangular area with $10 \lesssim t_{esc} \lesssim
 10^4$. This area becomes wider in $t$ with increasing
$C$: $10 \lesssim t_{esc} \lesssim 10^3$ corresponds to $C=0.01$ and $10 \lesssim
t_{esc} \lesssim 10^4$ corresponds to $C=0.10$. These are the particles
that are associated with the initial fast decay. The widening of
this area can be explained with our model from section
\ref{ch:model}. In the case $C=0.01$ (inset of Fig.
\ref{fig:EscTime_vs_F_Amp_001_IV}) the particles with $F>0$
escape, similar to the static case with an exponential rate, and after a
certain time, e.g. $t_{esc} \approx 1000$, most of them are escaped. Consequently,
the algebraic decay establishes itself, see Fig.
\ref{fig:EscSL_IV}. With increasing  amplitude, horizontal and
vertical processes lead to larger changes $\triangle F$.   Due
to multiple separatrix-crossing scattering the available range in
the time $t$ to escape clearly becomes larger. The appearance of rather  high
densities  at $F\approx 0.9$ and
$F\approx 0$  in Fig. \ref{fig:EscTime_vs_F_Amp_001_IV} and at
$F\approx 1$ in the inset, is explained below, in section
\ref{ch:DistrF}.

For values of $F<0$, corresponding to librator orbits, the values
of the escape time are  grouped in an inclined band, i.e. for
smaller values of $F$, the escape time is on average higher. In
the case $C= 0.01$, the band stops at $F \approx -0.20$, orbits
with smaller initial values of $F$ just did not escape until
$5\cdot 10^4$ collisions were reached.  For $C=0.10$, this  band
covers almost the hole range in F,  due to the larger driving
amplitude and the  larger effective changes $\triangle
F$.

In Fig. \ref{fig:EscTime_vs_F_Amp_005_HV}, the results of the
HVE are shown. We can match perfectly the exponential short-time
behavior and the algebraic tail with the two major areas in
picture.  All particles with initial values  $F>0$,
corresponding to rotator orbits, have escape times of $t_{esc}<10$,
whereas particles on librator orbits with initial values of $F<0$ possess
escape
times $t_{esc}>10$. Due to high particle velocities,
the effects of the horizontal and vertical processes are rather
small, it takes around $t\approx 10$ (corresponds to approximately $500$
collisions) until the first librators are destroyed. Up to $2.5 \times
10^4$ collisions, only
particles with $F\gtrsim 0$ escaped ($C=0.05$), whereas for
$C=0.25$ particles with $F\gtrsim -0.5$ decayed.

For values $F<0$ and $t_{esc}>10$  horizontal, narrow layers can
be observed  in Fig. \ref{fig:EscTime_vs_F_Amp_005_HV}. The
vertical spacing of these layers is  $2\pi$, which  is again the
period of the breathing ellipse. The mechanism at work is the
previously (section \ref{ch:TDEscapeRate}) mentioned one: When the
ellipse is expanding, librators are turned into rotators, which
can then escape, whereas during the contraction period, the
rotators are stabilized.
\begin{figure}[ht]
        \includegraphics[width=0.9\columnwidth]{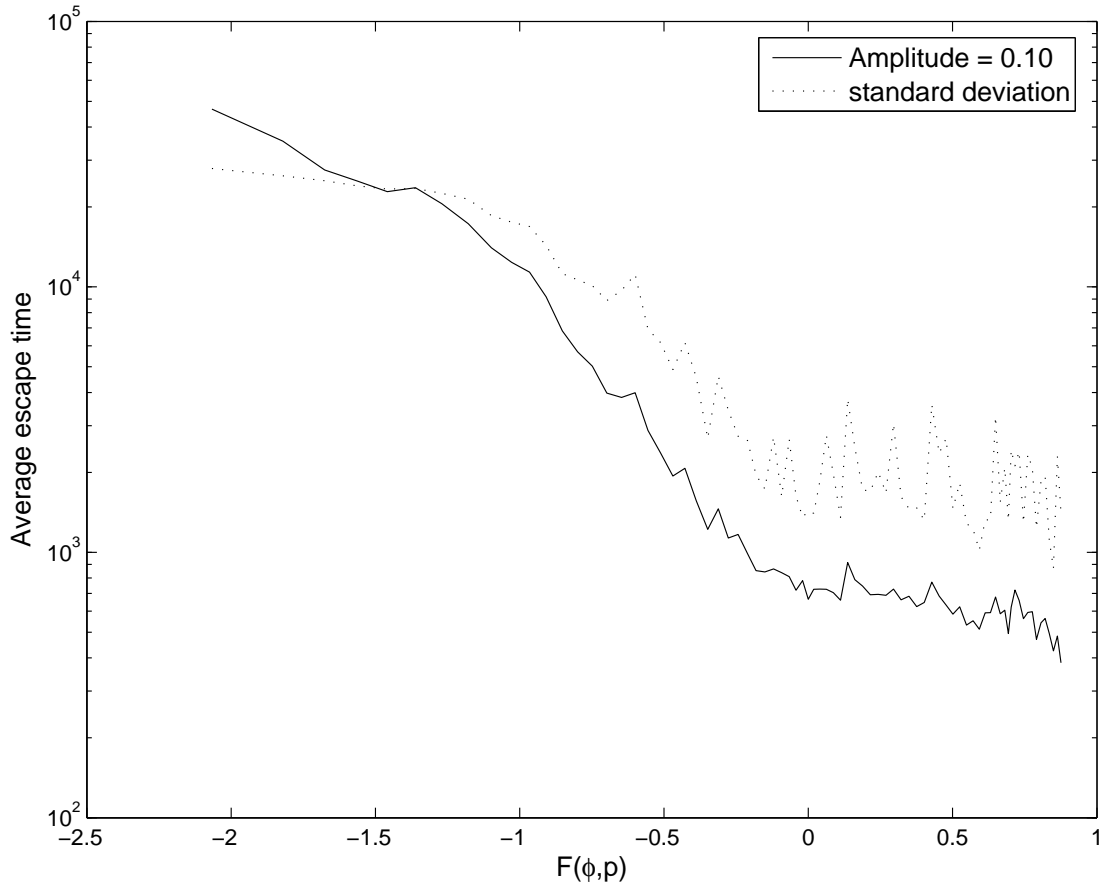}
        \caption{Average escape time and standard deviation in the IVE as
a function of the initial value of  $F(\varphi_0,p_0)$ for $C=0.10$.
}\label{fig:AvEscTime_vs_F_Amp_010_IV}
\end{figure}
Exemplarily, the average escape time as a function of the initial
value of $F$ is shown in Fig.
\ref{fig:AvEscTime_vs_F_Amp_010_IV} for $C=0.10$ and the IVE. Starting from
$F$ around $-2$, the escape time is decreasing with increasing
$F$, until the separatrix ($F=0$) is reached. For values of $F$
bigger than zero, corresponding to rotators, the escape time
stays approximately constant. Nevertheless, the escape time for a
single trajectory with a certain initial value $F$ can deviate
significantly form the curve shown in Fig.
\ref{fig:AvEscTime_vs_F_Amp_010_IV}, since the standard
deviation, also shown in Fig. \ref{fig:AvEscTime_vs_F_Amp_010_IV}, is quiet
large, especially for values of $F>0$.

\subsection{Density Distributions of the Escaping/Nonescaping
PAM}\label{ch:DistrF}
\begin{figure}[ht]
        \includegraphics[width=0.9\columnwidth]{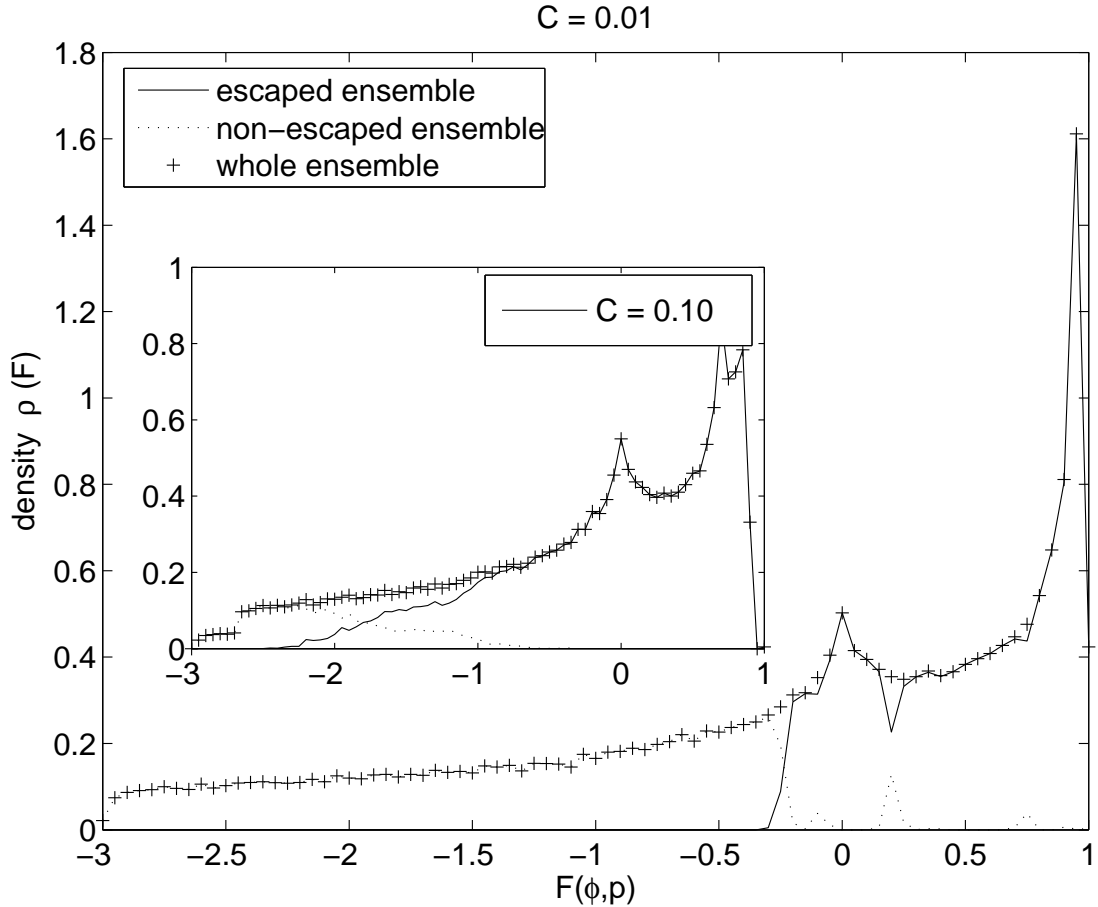}
        \caption{Density distribution of  $F(\varphi_0,p_0)$ in the IVE of the
whole, the escaped and the remaining ensemble for $C=0.01$ and $C=0.10$ (inset).
}\label{fig:F_DensityEscRem_Amp_001_IV}
\end{figure}

\begin{figure}[ht]
        \includegraphics[width=0.9\columnwidth]{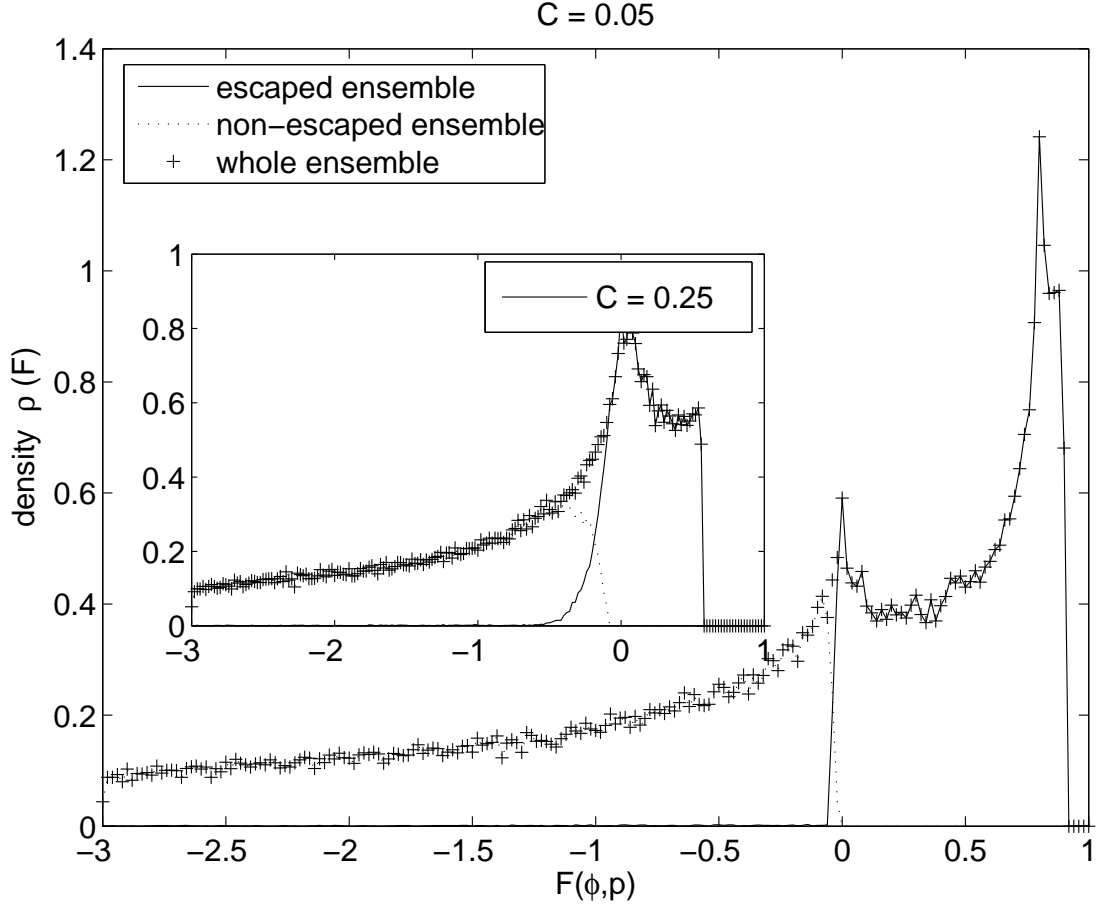}
        \caption{Same as Fig. \ref{fig:F_DensityEscRem_Amp_001_IV} for the
HVE, $C=0.05$ and $C=0.25$ (inset). }\label{fig:F_DensityEscRem_Amp_005_HV}
\end{figure}
The choice of the initial conditions described in section
\ref{ch:TDEscapeRate} leads to an amplitude-dependent, non-uniform
distribution in the density $\rho(F(\varphi_1,p_1))$ after the first collision
at
$(\varphi_1,\,p_1)$. In Fig.
\ref{fig:F_DensityEscRem_Amp_001_IV}, the density distribution of the
initial values of $F(\varphi,p)$ for $C=0.01 \text{ and } 0.10$ is
shown (IVE). The three different curves in each of the figures
correspond to the initial ($t=0$) \cite{endnote37} density distribution of
$F(\varphi,p)$ of the
whole ensemble, to the initial density distribution of $F$ for the escaped and
to the density distribution of $F$ for the remaining (i.e. non-escaped)
particles:
\begin{equation}
\rho_{all}(F_0) =  \rho_{rem}(F_0) + \rho_{esc}(F_0), \qquad
\int_{F_{min}}^{F_{max}} \rho(F_0)dF_0=1.
\end{equation}
If we compare these figures with
Fig.
\ref{fig:EscTime_vs_F_Amp_001_IV}, there is a perfect
correspondence between the peaks of the initial density distribution of
$F(\varphi,p)$ and the high density regions in the latter Fig.,
i.e. the high density regions are due to the non-uniformity of the
distribution of the initial values of $F$.

With increasing amplitude, the available range of initial values
of $F$ still making to an escape of the particles possible becomes
larger, for $C=0.01$ only particles with $F_0\gtrsim -0.3$
escaped, whereas for $C=0.10$ already particles with initial $F_0\gtrsim -2.4$
escaped.

The results of the HVE are very similar, see Fig.
\ref{fig:EscTime_vs_F_Amp_005_HV}. The main difference is, that
due to the small effect of a single scattering process, the
transition between escaping and nonescaping particles in $F$-space is much
sharper and shifted towards higher values
of $F$ compared to the IVE. Furthermore, the density  distribution of the
initial values of $F$ explains the reverse ordering of
$N_C(t=50)$ observed in section \ref{ch:TDEscapeRate}. For
$C=0.05$, due to our definition of the initial ensemble at the innermost ellipse
boundary, there are much more particles with initial values
$F\lessapprox 1$, than in the case $C=0.25$. Overall, the fraction of particles
starting on rotator orbits in the case $C=0.05$ is larger than
in the case $0.25$, and these rotators will escape
fast.

\section{Velocity}\label{ch:Velocity}
In the static ellipse, see section \ref{ch:StaticEllipse}, there
are two constants of motion. One is  the product of the angular
momenta around the two focus points $F(\varphi,p)$, which we just
studied in the context of the driven ellipse in the previous
section, the other is the energy. Since the potential is constant
inside the ellipse, it is sufficient to consider the kinetic
energy only. The  energy of a single particle in the ellipse is
given by $E_{total} = E_{kin} =  m \bm{v}^2/2$. Energy
conservation in the static ellipse thus means $|\bm{v}|=const$.
Since all particles have the same mass it is sufficient to
consider $|\bm{v}|$ instead of $E_{total}$.

When examining $F(\varphi,p)$, it is instructive to calculate
$F$ from the initial conditions $(\varphi_0,\,p_0)$.  Doing the
same in the  case of the velocity $|\bm{v}(t)|$ is
meaningless, since we  know $|\bm{v}(0)|=1$ (IVE) or
$|\bm{v}(0)|=100$ (HVE) for all particles. Instead, we consider
the velocity of the particles when they are actually escaping,
i.e. $|\bm{v}(t_{esc})|$, the \textit{escape velocity}.

\subsection{Escape Velocity versus Escape Time}\label{ch:CorTimeVel}

\begin{figure}[ht]
        \includegraphics[width=0.9\columnwidth]{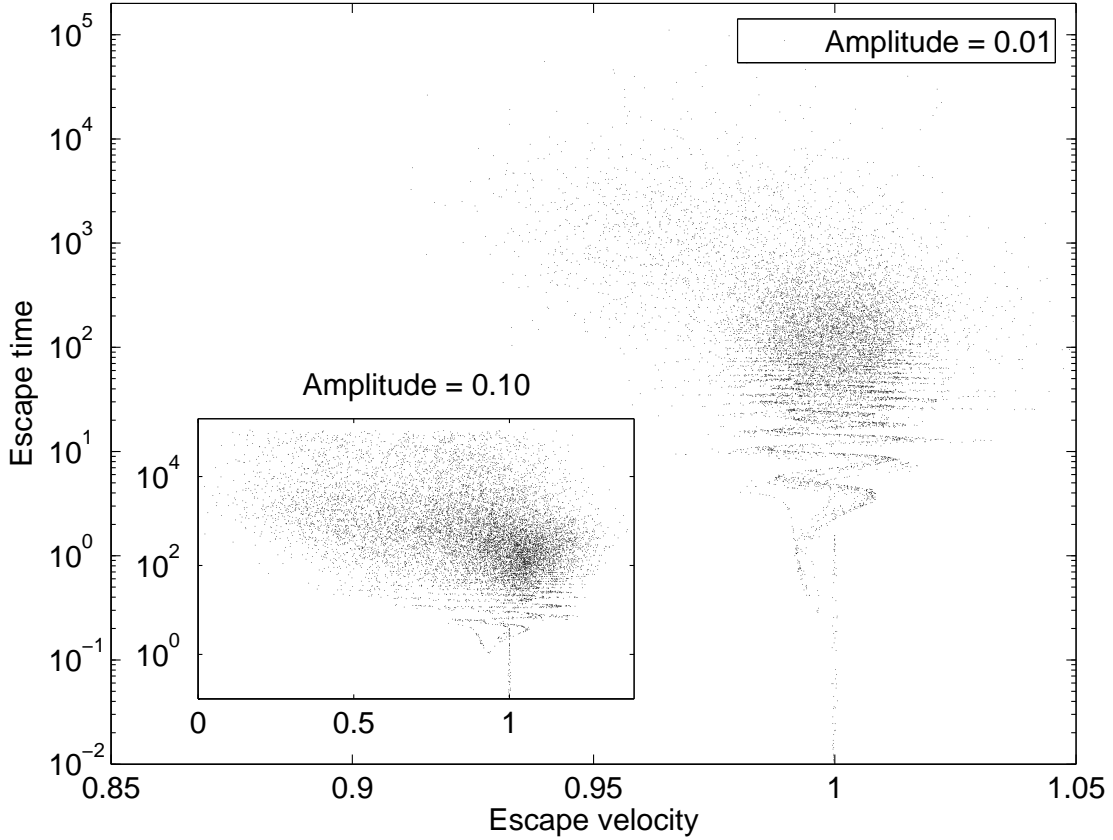}
        \caption{Escape time versus the absolute value of the escape velocity in
the IVE  for $C=0.01$ and $C=010$ (inset), each point corresponds to a pair
$(|\bm{v}(t_{esc})|,\,t_{esc})$. }\label{fig:EscTime_vs_V_Amp_001_IV}
\end{figure}
\begin{figure}[ht]
        \includegraphics[width=0.9\columnwidth]{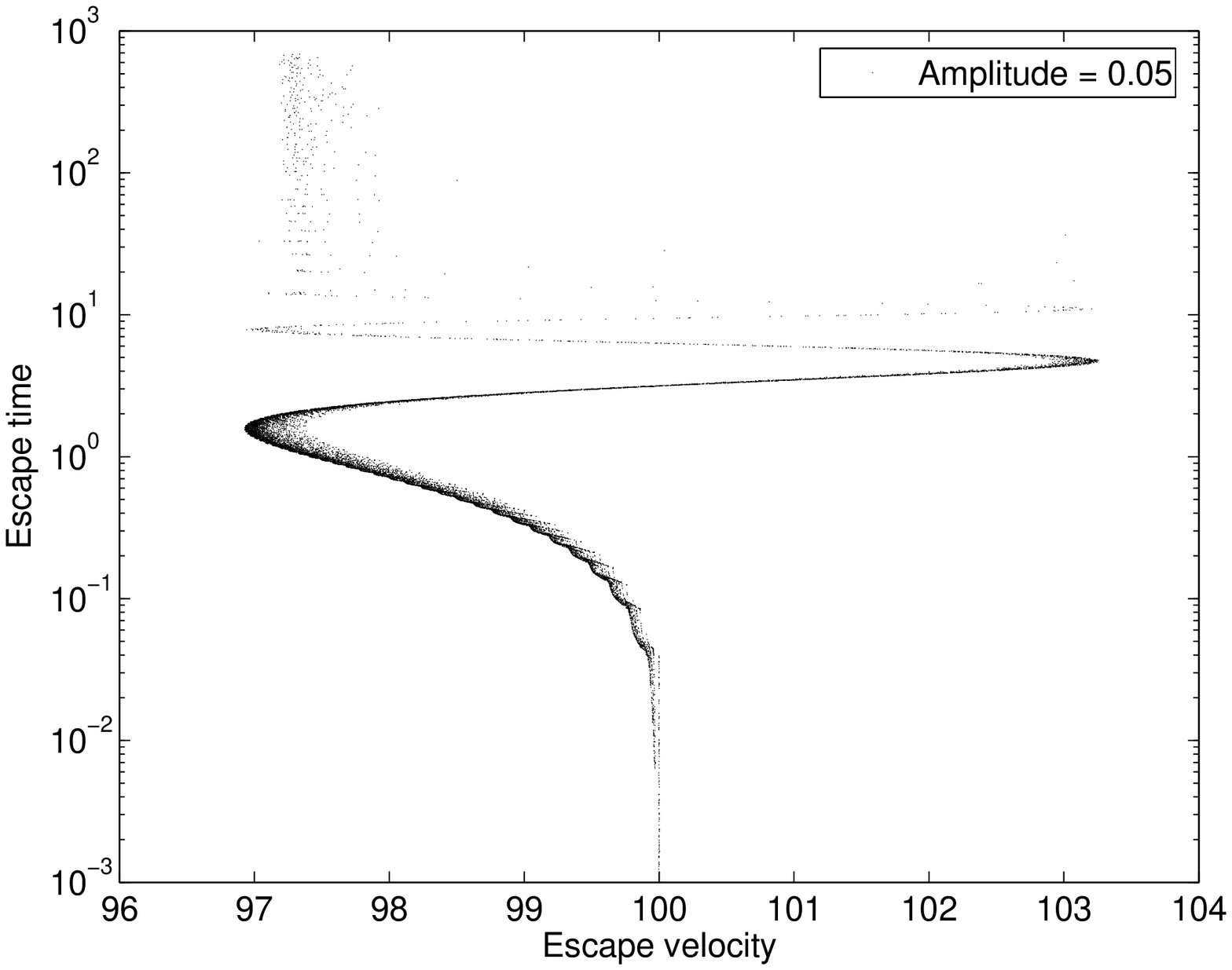}
        \caption{Same as Fig. \ref{fig:EscTime_vs_V_Amp_001_IV} for the HVE,
$C=0.05$. }\label{fig:EscTime_vs_V_Amp_005_HV}
\end{figure}
In Fig. \ref{fig:EscTime_vs_V_Amp_001_IV} distributions of escape times as
functions of the escape velocity
are shown for $C=0.01$ and $C=0.10$. A point
for every pair $(|\bm{v}(t_{esc})|,\,t_{esc})$ is plotted in the
plane and only escaped particles are considered (IVE). At
$|\bm{v}|=1$, there is a vertical line for $t \lesssim 4$. This
line corresponds to the particles that escape from the billiard
without a single boundary collision and possess thus an unchanged
energy. For times  $t \lesssim 10^2$, the pairs
$(|\bm{v}(t_{esc})|,\,t_{esc})$ lie on a narrow serpentine band.
The vertical spacing, i.e. the period of the band is approximately
$2\pi$, which is the period of the driven ellipse. The band
structure is much more pronounced in the case of the HVE, see
Fig. \ref{fig:EscTime_vs_V_Amp_005_HV}, where it dominates the
overall distribution,  naturally emanating from $|\bm{v}|=100$. This
correlation between the escape time and the escape velocity for
small values of $t_{esc}$ can be explained in the following way.
The ellipse starts at $t=0$ from its neutral position with an
expanding motion. As long as the ellipse is expanding, each time a
particle hits the boundary it loses energy and its velocity is
reduced. Since the particles move very fast compared to the motion
of the boundary (HVE), they accumulate a lot of collisions until
the ellipse reaches its maximal extension and starts contracting.
The more collisions a particle cumulates during the expansion
period, the bigger is the total energy loss. The ellipse reaches
its turning point at $t=\pi/2$, i.e. every particle with an
escape time $t_{esc}\leq\pi/2$, will have an escape velocity
$|\bm{v}(t_{esc})|\leq|\bm{v}_0|=100$. From $t=\pi/2 $ on, the
corresponding escape velocities will increase until
$t_{esc}=3\pi/2$ is reached, since the ellipse is contracting
during this time period and every collision with the boundary will increase
the energy of the reflected particle. This process is continued
until all rotators have escaped, which is the case at $t\approx10$.
This explanation holds for the HVE. Since in the case of the IVE,
the particle velocities are similar to the boundary velocity, this
effect is much less pronounced. Nevertheless, it is
still visible and mainly due to orbits with initial values
$F\approx 1$, since these orbits skip along the ellipse,
accumulating many boundary collisions within a short period of time.
The main difference of the distributions of the escape time  for the two
above-investigated ensembles is that in the case of the HVE all rotator orbits
lie on the serpentine band, whereas in the case of the IVE, only rotators
far away ($F\approx 1$) from the separatrix  contribute.

For intermediate times $10^2 \lesssim t_{esc} \lesssim 10^3$
(IVE),  the corresponding escape velocities lie closely around one
for $C=0.01$, see Fig. \ref{fig:EscTime_vs_V_Amp_001_IV}. Since
the driving amplitude is very small in this case, the energy of
the particles is not much changed. For larger escape times, the
values of $|\bm{v}(t_{esc})|$ are a little bit more scattered,
since the particles accumulated several boundary collisions
resulting in an effective change of $|\bm{v}(t_{esc})|$, still the
values deviate no more than $10\%$ from the the initial value
$|\bm{v}| =1$.  As expected, the distribution of the values of
$|\bm{v}(t_{esc})|$ is broadened for larger values of the driving
amplitude $C$, see inset of Fig. \ref{fig:EscTime_vs_V_Amp_001_IV}
($C=0.10$), but the energy gain remains bounded,
$|\bm{v}(t_{esc})| \lesssim 3$ even for $C=0.30$ (not shown here).

In the HVE, for escape times higher than 10, corresponding to
particles starting originally on librator orbits, almost all escape
velocities are  smaller than the initial velocity
$|\bm{v}_0|=100$.  Horizontal and vertical processes can scatter a
particle, moving on a librator orbit, onto a rotator orbit (and vice
versa), which is a necessary condition for escaping. In which
direction (towards or away from the elliptic fixed points) a
particle is scattered depends on many parameters, see section
\ref{ch:DesTypII}, but at least the vertical process scatters
particles during the expansion period of the ellipse \emph{always}
towards rotator orbits.  As a consequence, particles that turn
from librator into rotator orbits accumulate collisions that
effectively reduce their velocity and scattered them beyond the
separatrix, thereby explaining the low escape velocities of particles
with escape times $t_{esc}>10$. One might think of using this mechanism to slow
down particles.

\subsection{Distributions of the Escape Velocities}\label{ch:DistrEscVel}

\begin{figure}[ht]
        \includegraphics[width=0.9\columnwidth]{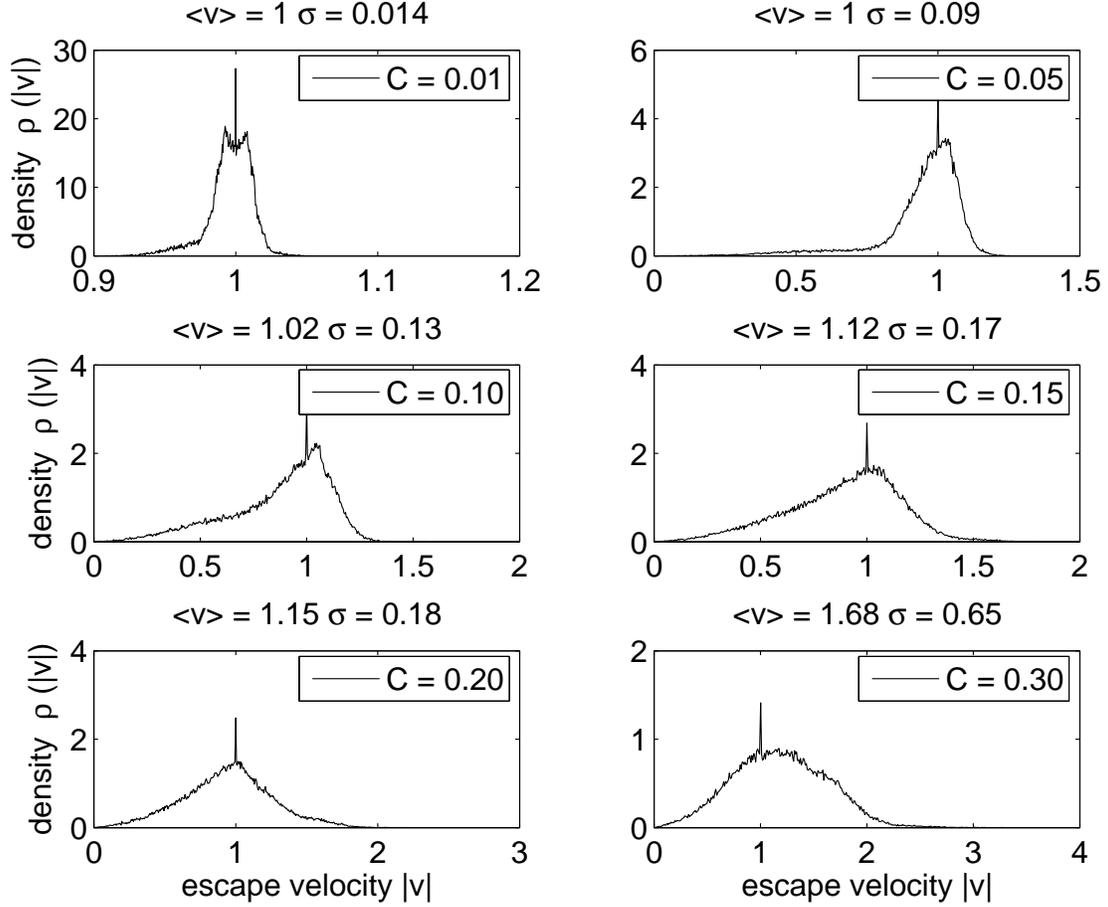}
        \caption{Distribution of the escape velocity in the IVE for different
values of $C$. }\label{fig:EscVelDistr_Esc_Many_Amp_IV}
\end{figure}
\begin{figure}[ht]
        \includegraphics[width=0.9\columnwidth]{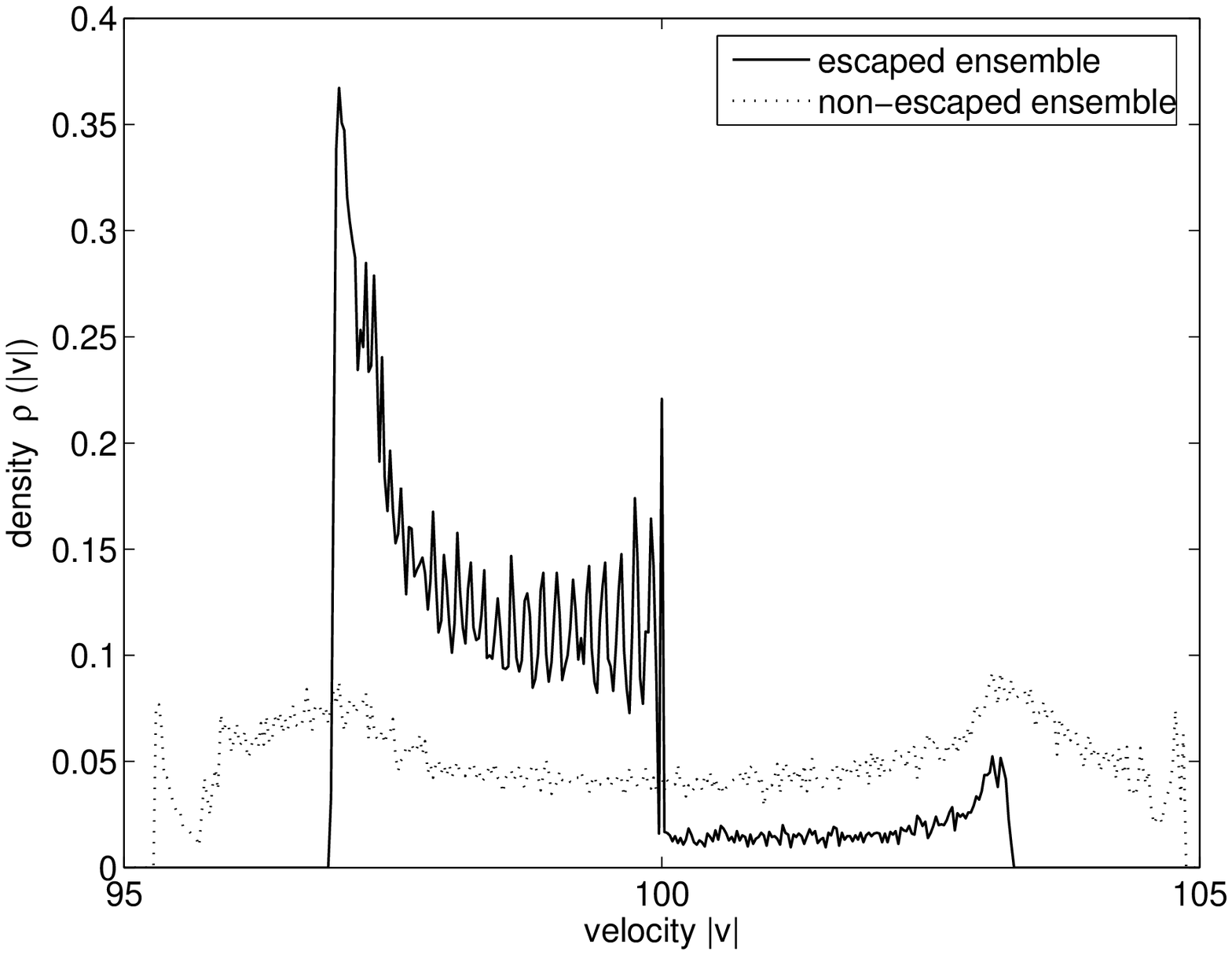}
        \caption{Distribution of the escape velocity for the HVE, $C=0.05$.
}\label{fig:EscVelDistr_EscRem_Amp_005_HV}
\end{figure}
In Fig. \ref{fig:EscVelDistr_Esc_Many_Amp_IV}, the distribution of
the escape velocities is shown for different values of $C$ (IVE).
With increasing amplitude, the mean escape velocity
$<|\bm{v}(t_{esc})|>$ is shifted towards larger values. Since on
average there are slightly more collisions with the contracting
ellipse than with the expanding one, a larger driving amplitude
leads to an increased mean energy of the particles. The sharp peak
at $|\bm{v}|=1$ corresponds to the particles that leave the
billiard without a single collision with the boundary, thus
$|\bm{v}_{esc}|=|\bm{v}_{0}|=1$.

In the case of the HVE, see Fig.
\ref{fig:EscVelDistr_EscRem_Amp_005_HV}, the distribution looks
quite differently.  The majority of the particles have an escape
velocity smaller than $|\bm{v}_0|=100$ and especially at the
lowest accessible escape velocity $|\bm{v}_{esc}|\approx 97$ there
is a large peak. This is due to the fact that the ellipse starts
with an expanding motion which deprives the particles energy upon
boundary collisions and a large fraction of particles decays
during that first expansion period. Around $t=\pi/2$, in the
vicinity of the first turning point, the ellipse stays
comparatively long (boundary velocity $\approx 0$), more particles
escape, leading to the large peak at $|\bm{v}|\approx 97$.  The
asymmetric shape of the distribution is additionally reinforced
since  the librators that escape have also low energies. The
particles that are not escaped after $2.5\cdot10^4$ collisions
accumulated  collisions during the expanding and contracting
motion of the ellipse and the fluctuations in the energy transfer
lead to a roughly uniform distribution. The distributions for
higher values of $C$ look very much like the one shown in the case
$C=0.05$, except that they get wider  with increasing amplitude.

Since  most of the particles leaving the ellipse have
velocities smaller than $|\bm{v}_0|$, the question arises, whether
the billiard could be used for systematic velocity lowering. To
enforce this effect, one could try e.g. to choose asymmetric
driving laws. We point out that the lowered energies of the escaping  particles is a feature of the dynamics of the  ellipse. In general (concerning other geometries) particles are more likely to strike a contracting than a receding boundary, which leads one to expect increased energies of the escaping particles.

\section{Concluding remarks}\label{ch:Summery}
We investigated the classical dynamics of the static and
especially the driven elliptical billiard with an emphasis on the
escape rate of an ensemble of particles. As predicted in ref.
\cite{Bau90} in a general context for integrable billiards, we
found an algebraic decay in the long-time behavior of the static
ellipse, due to the integrable structure of the underlying
dynamics.  Besides the energy, the product of the angular momenta
(PAM) $F(\varphi, p)$ about the two foci  is preserved. The sign
of the initial value of $F$ determines whether a particle moves on
a rotator or librator orbit and only the rotators are always (for
all hole positions) connected with the hole. Consequently, the
decay approaches a  saturation value $N_s(\varepsilon)$, which is
maximal for the hole lying at the short side of the ellipse; at
this hole position none of the librators are connected with it.
$N_s(\varepsilon)$ depends on the numerical eccentricity
$\varepsilon$ of the ellipse and we predicted this dependence very
accurately from theoretical considerations. As a consequence,
varying $\varepsilon$ allows us to control the number of emitted
particles.

When applying harmonic boundary oscillations, neither the energy
nor $F(\varphi,p)$ will remain a constant of the motion. We
performed numerical simulations for two different ensembles,
representing the two important borderline cases. Firstly, the
intermediate velocity ensemble (IVE), where $|\bm{v}_0| \approx
\omega A$ ($\omega A$ being the boundary velocity), and secondly
the high velocity ensemble (HVE), where  $|\bm{v}_0| \gg \omega
A$. In both cases we observed an initial fast decay with an
ensuing transition period followed by a non-vanishing (even for
large times) near algebraic decay. The emission rate depends
monotonically on the driving amplitude. The changes of $F(\varphi,
p)$ of  particles upon a single collision with the boundary are
much smaller in the case of the HVE, due to the high velocities of
the particles. As a consequence the resulting decay is similar to
the one of the static system. The observed disappearance of the
saturation value in both ensembles is due to the gradual
destruction of the librator orbits caused by two fundamental
processes: The vertical processes, where upon collisions momentum
normal to the boundary is transferred, making changes in the sign
of $F(\varphi, p)$ possible; and the horizontal processes where the
particle hits the ellipse due to the boundary motion at a
different position (compared to the static case) leading again to
changes in $F(\varphi,p)$ that can result in the transition of a
librator into a rotator. Just like in the static system,
particles starting on rotator orbits $(F(\varphi_0,p_0) < 0)$ cause
the initial fast decay. With increasing time,
more and more particles with initial conditions closer and closer
to the elliptic fixed points can escape, due to the just described vertical and
horizontal processes, and cause the non-vanishing emission rate in the
long-time behavior of the decay. We confirmed this just displayed strong
connection between the escape time and  $F(\varphi_0,p_0)$ by analyzing this
quantity carefully. In the HVE, the escape rate as well as correlations of the
escape time and the PAM are modulated with the same period as the ellipse
breathes, the ellipse acts as a pulsed source.

Concerning escape velocities, an astonishing feature is observed in the case
of the HVE, the
distribution of the escape velocities is highly asymmetric and
particles escape mainly with $|\bm{v}(t_{esc})| < |\bm{v}_0|$, the
driven ellipse could be used for systematic cooling. To avoid
escape velocities bigger than $|\bm{v}_0|$, the use of a point
source as an initial ensemble seems reasonable. Simulations with thermal
ensembles suggested the ellipse as a state transformer, thermal ensembles were
changed into non-thermal ones. Furthermore, the ellipse could be used as a
controlable source of particles: if a certain emission rate is
is required, this can be achieved by tuning the driving amplitude, whereas the
numerical eccentricity $\varepsilon$ of the static ellipse allows us to emit a
certain number of particles.

\section{Acknowledgments}\label{ch:acknowledgments}
Valuable discussions with A. Richter, V. Constantoudis, A. Karlis and M.
Oberthaler are gratefully acknowledged. F.L. acknowledges support from
the Landesgraduiertenf\"orderung Baden-W\"urttemberg. F.K.D. likes to thank the DAAD for financial support in the framework of a visit to the University of Heidelberg.

\end{document}